\documentclass[aps,prl,showpacs,notitlepage,nofootinbib,superscriptaddress,floatfix,showkeys,twocolumn,preprintnumbers]{revtex4-1}

\usepackage{blindtext}
\usepackage{hyperref}
\usepackage{amsmath,amssymb}
\usepackage{float}
\usepackage{microtype}
\usepackage{graphicx}
\usepackage{bm}
\usepackage{latexsym}
\usepackage{epsfig}
\usepackage{psfrag}
\usepackage{color}
\usepackage[dvipsnames]{xcolor}
\usepackage{subfigure}
\usepackage[section]{placeins}
\usepackage{comment}

\def\e1i{\epsilon_{1\mathrm{i}}}

\allowdisplaybreaks[1]

\begin{document}
\preprint{KCL-2024-52}

\title{Anomalous Ionization in the Central Molecular Zone by sub-GeV Dark Matter
}

\author{Pedro De la Torre Luque}\email{pedro.delatorre@uam.es}
\affiliation{Departamento de F\'{i}sica Te\'{o}rica, M-15, Universidad Aut\'{o}noma de Madrid, E-28049 Madrid, Spain}
\affiliation{Instituto de F\'{i}sica Te\'{o}rica UAM-CSIC, Universidad Aut\'{o}noma de Madrid, C/ Nicol\'{a}s Cabrera, 13-15, 28049 Madrid, Spain}
\affiliation{The Oskar Klein Centre, Department of Physics, Stockholm University, Stockholm 106 91, Sweden}
\author{Shyam Balaji}
\email{shyam.balaji@kcl.ac.uk}
\affiliation{Physics Department, King’s College London, Strand, London, WC2R 2LS, United Kingdom}

\author{Joseph Silk}
\email{silk@iap.fr}
\affiliation{Institut d’Astrophysique de Paris, UMR 7095 CNRS \& Sorbonne Universit\'{e}, 98 bis boulevard Arago, F-75014 Paris, France}
\affiliation{Department of Physics and Astronomy, The Johns Hopkins University, 3400 N. Charles	Street, Baltimore, MD 21218, U.S.A.}
\affiliation{Beecroft Institute for Particle Astrophysics and Cosmology, University of Oxford, Keble	Road, Oxford OX1 3RH, U.K.}

\smallskip
\begin{abstract}
We demonstrate that the anomalous ionization rate observed in the Central Molecular Zone can be attributed to MeV dark matter annihilations into $e^+e^-$ pairs for galactic dark matter profiles with slopes $\gamma>1$. The low annihilation cross-sections required avoid cosmological constraints and imply no detectable inverse Compton, bremsstrahlung or synchrotron emission in radio, X and $\gamma$ rays. The possible connection to the source of the unexplained $511$ keV line emission in the Galactic Center suggests that both observations could be correlated and have a common origin.
\end{abstract}
\maketitle

\begin{figure}[t!]
\includegraphics[width=\linewidth]{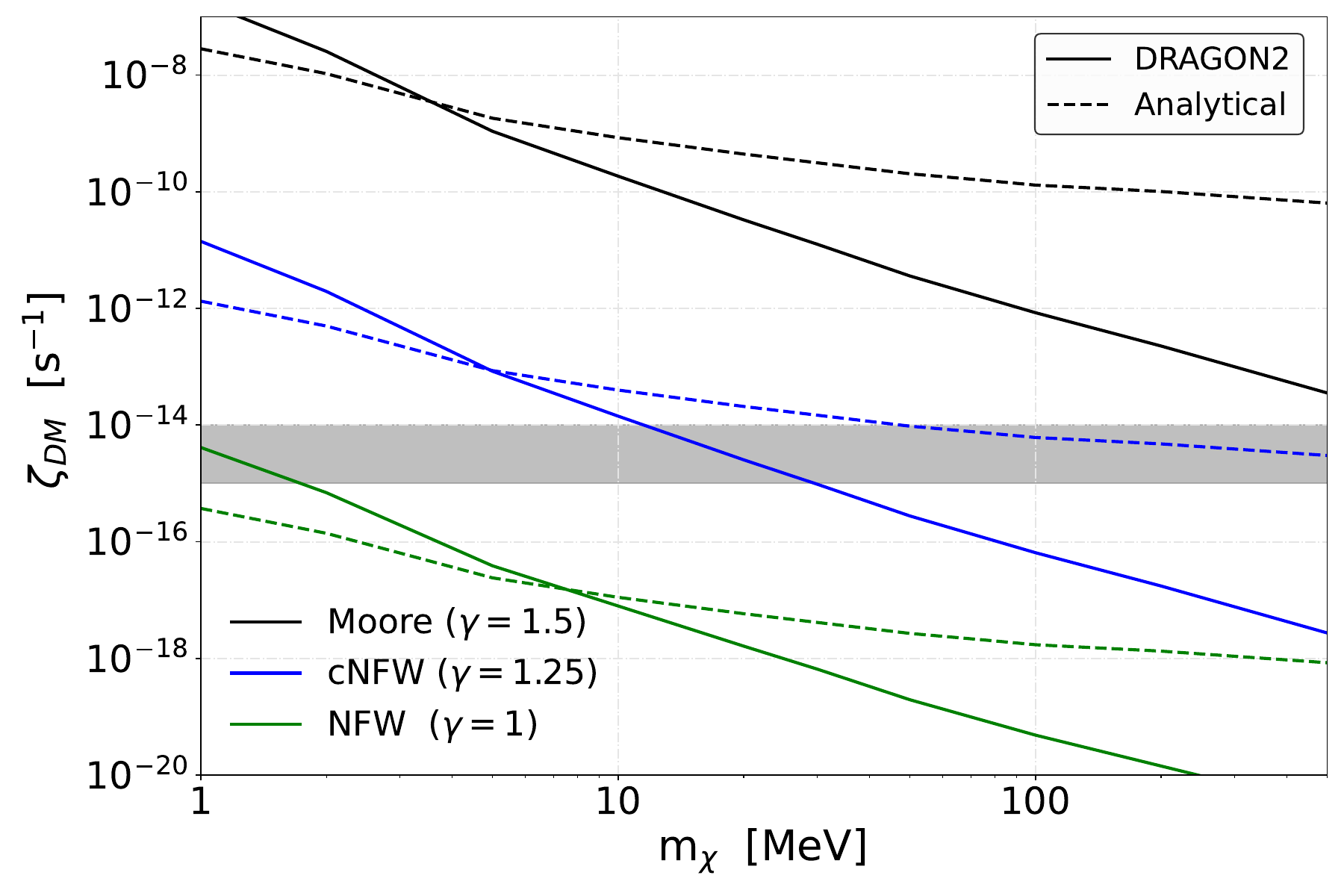}
\includegraphics[width=\linewidth]{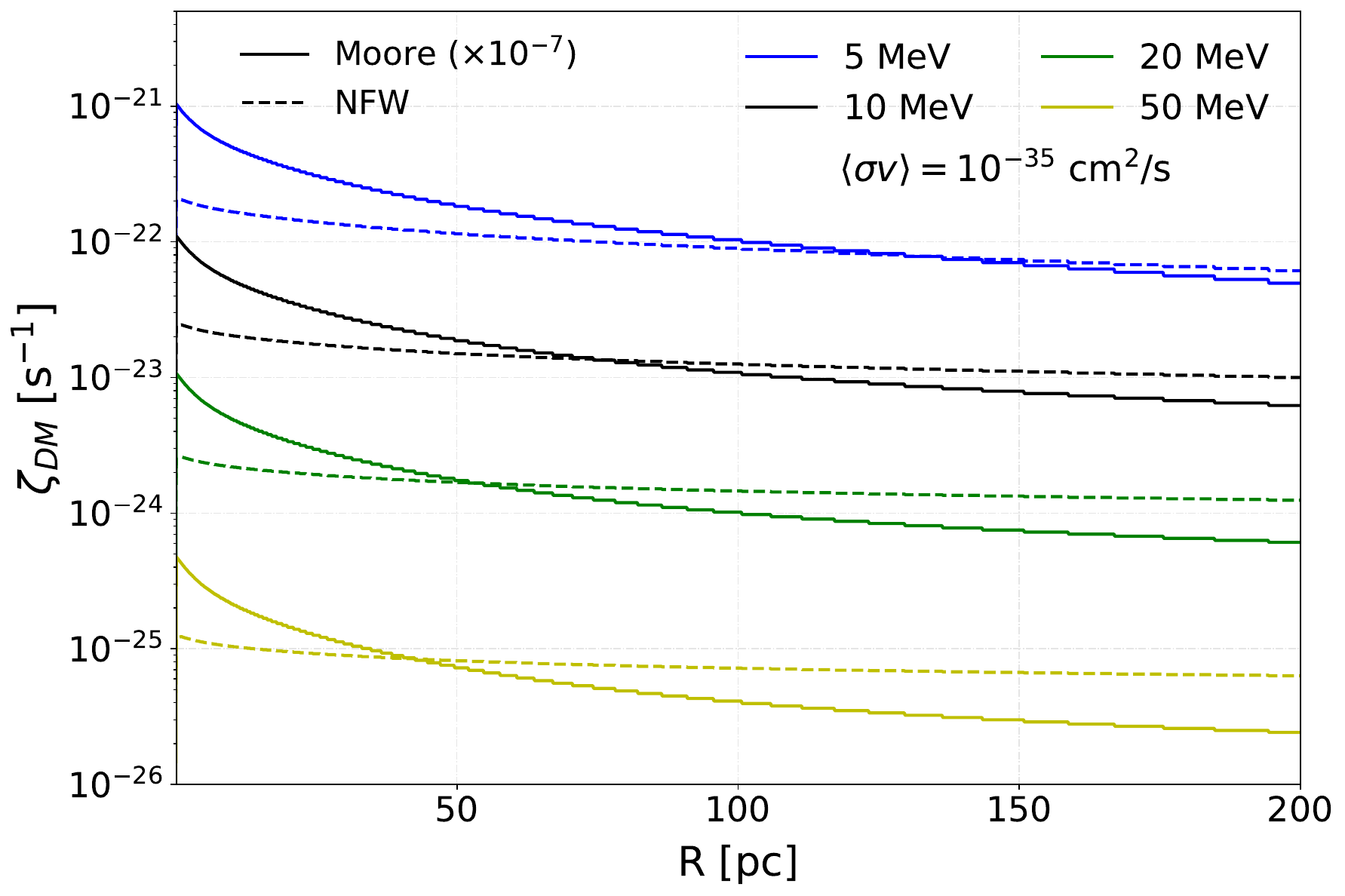}
\caption{\textbf{Top panel}: Maximum ionization rate that can be provided by sub-GeV DM (i.e. the average ionization rate within the CMZ obtained from the upper limits on $\langle \sigma v \rangle$ constraints allowed by CMB~\cite{Lopez_Honorez_2013, Slatyer_2016}).
We compare the predicted ionization rate obtained with an analytical calculation and with a numerical calculation using a dedicated CR propagation code ({\tt DRAGON2})~\cite{DRAGON2-1, DRAGON2-2, de_la_torre_luque_2023_10076728}. The shaded area represents the ionization rate observed in the CMZ~\cite{IoRCR, RevModPhys.92.035003, Indriolo_2015, 2022FrASS...9.9288R, Yusef_Zadeh_2012, 2007ApJ...656..847Y, Oka_2005, Goto_2011, Geballe_2010, IoRPetit, Sanz-Novo_2024}.
\textbf{Bottom panel}: Radial profile of the ionization rate expected from DM annihilation ($\langle \sigma v \rangle = 10^{-35}$~cm$^3$/s) for the Moore (solid line, multiplied by $10^{-7}$ to have an easy comparison) and NFW (dashed lines) profiles. The different colors indicate different DM masses. }
\label{fig:IonRate}
\end{figure}

\emph{Introduction---}
Various measurements are known to indicate an unexpectedly high rate of H$_2$ ionization rates around the Galactic Center (GC), particularly in the Central Molecular Zone (CMZ). This has been evidenced by a variety of tracers, including methanol production~\cite{IoRMethanol}, (indicating an ionization rate of $\zeta \sim 10^{-15}$~s$^{-1}$), and the observation of $\textrm{H}_3^+$ lines~\cite{Oka_2020, Indriolo_2009, Goto_2014} (requiring an ionization rate around $\zeta \sim 2\times10^{-14}$~s$^{-1}$)~\cite{IoRPetit, IoRCR, Geballe_2010} and other ionized molecules, such as $\textrm{H}_3$O$^+$~\cite{Indriolo_2015, van_der_Tak_2006} (indicative of an ionization rate $\gtrsim 10^{-15}$~s$^{-1}$).

Recently, Ref.~\cite{IoRCR} showed that the diffuse flux of cosmic rays (CRs) falls short of explaining such high ionization rates in the CMZ, arguing that it requires an extremely large injection power for CRs at the GC, of around $10^{40} $-$ 10^{41}$~erg s$^{-1}$ (see also Ref.~\cite{Phan_2018}). 
This may indicate the presence of a source highly concentrated around the GC that is producing low energy CRs that lose their energy very quickly, via ionization of molecular gas~\cite{Indriolo_2009}. From the latter, one can infer that these particles must be injected at low energies, to be efficiently confined around the CMZ thereby preventing them from propagating long distances.

Here, we investigate whether annihilation of sub-GeV DM \cite{Boudaud:2016mos, Cirelli:2023tnx} into $e^{\pm}$ pairs can explain the observed ionization rate for DM masses below $\sim100$~MeV, without being in conflict with other known current constraints. While DM density distributions with a central spike could also provide promising results, in this work we consider several dark matter profiles including the Navarro-Frenk-White (NFW)~\cite{Cirelli:2024ssz} and those with larger slope $\gamma$.

The motivation for sub-GeV DM is strong and has previously been discussed in detail in the literature, see e.g. Refs.~\cite{Battaglieri:2017aum,Cirelli:2020bpc,Balan:2024cmq} and Ref.~\cite{Cirelli:2024ssz} for a recent review.
Presently, CMB constraints~\cite{Lopez_Honorez_2013, Slatyer_2016}, provide the most robust constraints for annihilation of sub-GeV DM particles, and, as shown in Fig.~\ref{fig:IonRate}, these are consistent with a DM origin for the anomalous ionization rate observed in the CMZ for profiles generally cuspier than a NFW DM profile~\cite{Navarro:1995iw} and low DM masses (and even allow for combinations of mass and DM profiles that could exceed the observed ionization rate), due to the fact that the (low-energy) DM products can deposit roughly all their energy within the CMZ.
Here we investigate whether Galactic probes are able to constrain and rule out this explanation of the CMZ ionization rate . 

The $511$~keV line emission from the bulge of the Galaxy remains a mystery~\cite{Prantzos:2010wi, kierans2019positron, Siegert_2023} and sub-GeV DM was proposed as a possible explanation~\cite{Boehm}. The emission can be used to strongly constrain sub-GeV DM, which injects Standard Model particles that yield electrons and positrons, forming positronium that can decay in the Galaxy~\cite{DelaTorreLuque:2023cef, Vincent_2012}. The observation of the line at the bulge requires a steady production of positrons at a rate of a few times $10^{50}$~$e^+$/yr~\cite{Siegert:2015knp, Kierans:2019aqz}. Once the electrons and positrons are produced, they lose energy, mainly via ionization of the molecular gas in the CMZ. If the medium is dense enough, as is the case for the CMZ, they release all their energy into the medium after travelling less than $\sim 100$~pc (for $e^{\pm}$ with energies below a few tens of MeV for standard diffusion parameters found in CR analyses). 
Once thermalized, the relativistic positrons produce positronium bound states, in the form of a singlet state, called para-positronium (p-ps), that decays into two $\gamma$ rays each with $511$~keV energy, or in a triplet state, called ortho-positronium (o-ps), that decays into $3$ photons~\cite{Badertscher_2007} producing continuum emission below $511$~keV.
We will also discuss the possibility that both observations are connected and explore whether the parameter space of sub-GeV DM that can explain both effects is currently ruled out. 

The remainder of this paper is as follows: we first detail our estimates of the CMZ ionization rate from the annihilation of sub-GeV DM particles. Then, we test whether the observations of the CMZ ionization rate are compatible with cosmological constraints and if there would be any associated significant inverse Compton or bremsstrahlung emission. Finally, we discuss whether the CMZ ionization rate and the $511$~keV line emission, could share a common (DM annihilation) origin and comment on the potential of finding correlations between these two observables.

\emph{Ionization rate from sub-GeV DM annihilation in the CMZ --}
The CMZ, located around the GC, is an approximately cylindrical region around $200$~pc wide and $100$~pc high~\cite{ferriere2007spatial, IoRCR}, of high-density molecular gas ($n_\textrm{H}\sim 150$~cm$^{-3}$). Low energy CRs $\lesssim 100$~MeV are subject to severe ionization energy losses that limit their propagation (see Fig.~S1 in Supplementary material).
At higher energies, these particles can escape the CMZ and produce potentially detectable secondary diffuse emission (such as bremsstrahlung, inverse Compton or synchrotron radiation).

The ionization rate $\zeta$ caused by the electrons and positrons of kinetic energy $E$ at position vector $\textbf{x}$ inside the CMZ can be calculated using Eq.~\eqref{eq:main} below~\cite{Padovani_2009, Phan_2023}.
\begin{equation}
  \zeta = 2\cdot 4\pi \int^{E_\textrm{max}}_{E_\textrm{min}} J(E, \textbf{x})\sigma(E)(1 + \theta_e(E)) dE 
   \,,
\label{eq:main}
\end{equation}
where the factor $2$ accounts for ionization by both $e^-$ and $e^+$, $E_\textrm{max}$ is the maximum kinetic energy with which the electrons or positrons are injected and $E_\textrm{min}$ is the minimum kinetic energy to produce ionization in the gas ($15.43$~eV~\cite{Padovani_2009, Phan_2023}), $\sigma(E)$ is the $e$-H$_2$ cross-section, calculated as in Eq.~(7) of Ref.~\cite{Padovani_2009}, $\theta_e(E)$ is the number of secondary ionizations per primary ionization, calculated as Eq.~(2.23) of Ref.~\cite{krause2015crimecosmicray}, and $J(E,\textbf{x})$ is the flux that can be computed as
\begin{equation}
    J(E, \textbf{x}) = \frac{d\phi_e}{dE}\left(E, \textbf{x} \right) \frac{\beta_e(E)c}{4\pi}
     \,,
\label{eq:flux}
\end{equation}
where $\beta(E) c$ is the speed of the electron or positron and $\frac{d\phi_e}{dE}$ is their differential flux in the CMZ. 
We compute the differential flux of electrons and positrons induced by sub-GeV DM annihilation in two ways: (i) an analytical approximation that allows us to estimate the diffuse $e^{\pm}$ flux within the CMZ, and (ii) a numerical calculation that allows us to compute their all-sky diffuse emission. We notice that both approaches used to compute the ionization rate are approximate, given the difficulties of treating diffusion in the CMZ, however, as we show below, the parameter space accessible to explain the measured ionization rate is so large that our conclusion should remain robust. We refer the reader to Section S.I of the Supp. Material for details on these calculations, which includes Refs.~\cite{DelaTorreLuque:2023huu, DelaTorreLuque:2023nhh, DelaTorreLuque:2023olp, DelaTorreLuque:2023cef, Ginz&Syr, Crank_Nicolson_1947, EventHorizonTelescope:2022wkp, AGUILAR20211, aguilar2018observation, AMS_gen, VoyagerMO, Stone150, Evoli_2020, Cirelli:2010xx}.
We also note that these annihilations are expected to also produce ionizing photons through final state radiation, however, the cross-section for ionizing H$_2$ is $\sim10^{-6}$ times lower than from $e^\pm$ at $1$~MeV.

\vspace{0.5cm}
\emph{Results ---}
The main constraints on sub-GeV DM arise from cosmological observations of the early Universe such as the effective number of relativistic species $N_\textrm{eff}$ and the requirement for successful Big Bang Nucleosynthesis (BBN). These observations can robustly rule out DM masses below around $1$~MeV \cite{Boehm:2012gr, Boehm:2013jpa, Nollett:2014lwa}, and even slightly above, depending on certain assumptions. However, recent work, such as Refs.~\cite{Escudero:2018mvt, Sabti:2019mhn}, has shown compatibility with observation for DM masses above a few MeV, as long as the dominant early Universe channel for DM annihilation is into neutrinos. 
For higher masses, there are strong astrophysical and cosmological constraints (see a recent review Ref.~\cite{balan2024resonantasymmetricstatussubgev} or a more comprehensive review of DM candidates Ref.~\cite{Cirelli:2024ssz}).

\begin{figure}[h!]
\includegraphics[width=\linewidth]{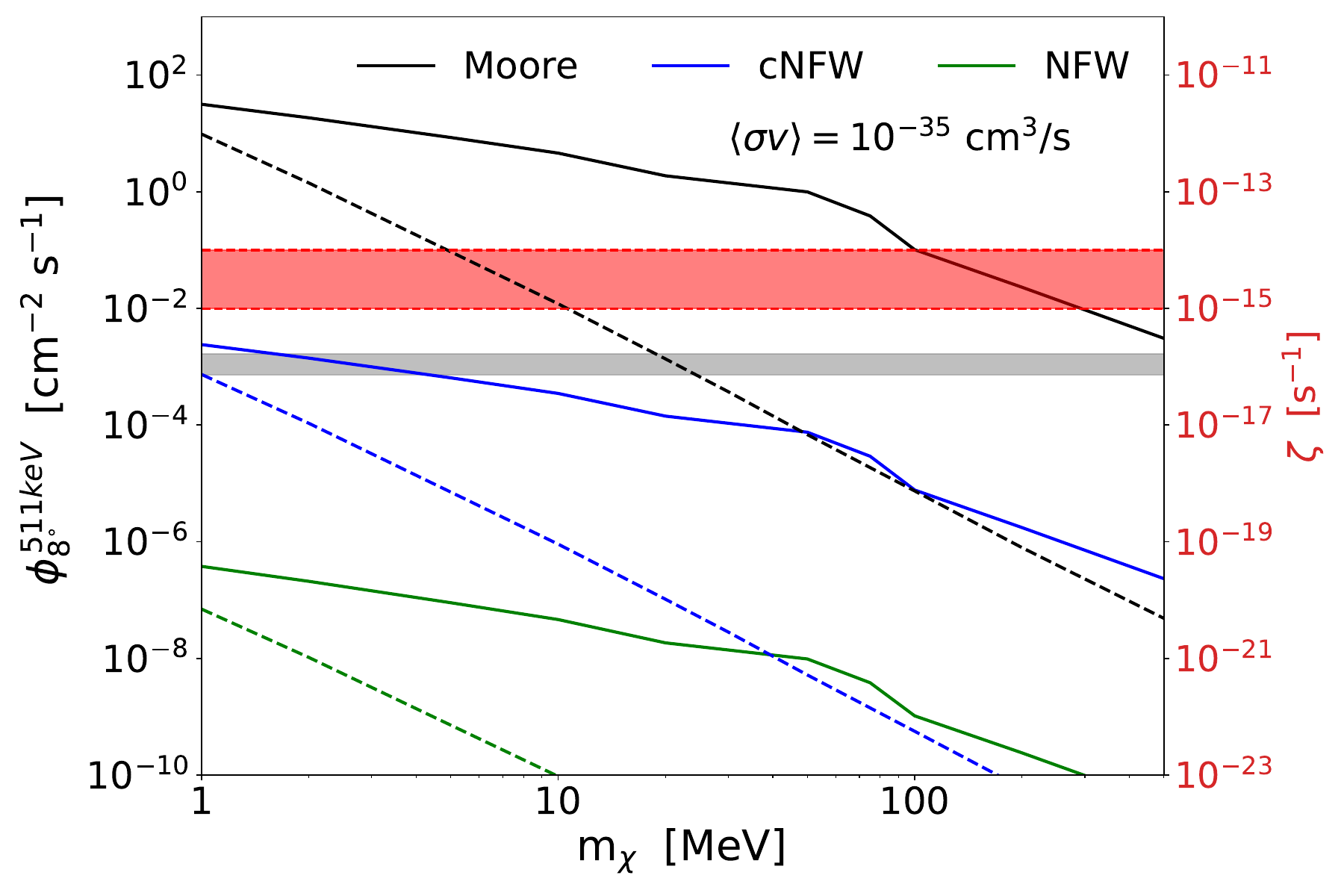}
\caption{\textbf{Left axis}: Predicted $511$~keV bulge emission associated with the $e^{\pm}$ population from annihilations for various DM density profiles (solid lines). The shaded band shows the constraint set by the observed emission from the bulge, with conservative $50\%$ uncertainties. \textbf{Right axis}: Expected ionization rate in the CMZ (i.e. averaged within the CMZ) from annihilation of sub-GeV DM particles into $e^+e^-$, as obtained with the numerical calculation (dashed lines). The red band illustrates the range of observed values of ionization rate in the CMZ.}
\label{fig:Lims}
\end{figure}

The estimated maximum ionization rate from sub-GeV DM with mass and cross-section combinations that are allowed by CMB constraints (obtained from the impact of DM annihilation on CMB anisotropies, see bottom left panel of Fig.~4 of Ref.~\cite{Slatyer_2016}), is shown in the top panel of Fig.~\ref{fig:IonRate}, where we represent in the $y$-axis the predicted ionization rate from sub-GeV DM in the CMZ for generalized NFW profiles with a slope ranging from $\gamma=1$ (standard NFW distribution), to $\gamma=1.5$ (a Moore profile~\cite{Moore_1999}), including what is known as a contracted-NFW (cNFW) profile, which corresponds to a value of $\gamma=1.25$, which is known for being the best-fit profile for the GeV Fermi-LAT Galactic Center Excess (GCE)~\cite{Ackermann_2017, Di_Mauro_2021}. 
In this figure, the shaded region represents the band of ionization rate values measured at the CMZ. Dashed lines represent the predicted ionization rate for the analytical approximation, while the solid lines represent the result obtained with the numerical calculation using {\tt DRAGON2}. The latter should be more representative, and we have checked that uncertainties from the diffusion parameters can affect the benchmark prediction by at most a factor of a few (see Ref.~\cite{DelaTorreLuque:2023olp}). 
As a consequence, our result shows that there is a large region of parameter space (of DM profile slopes and DM masses) that can explain the observed ionization rate without violating cosmological constraints. 
To provide some quantitative examples, consider a $\simeq 10$~MeV DM particle annihilating at a cross-section of 
$\langle \sigma v \rangle\simeq 5\times10^{-30}$ cm$^3\,\rm s^{-1}$, 
which can produce an ionization rate of $\zeta\simeq 2\times10^{-14}$~s$^{-1}$ for a slope of $\gamma=1.25$. For the same slope, a $1$~MeV DM particle could produce the observed ionization rate for $\langle \sigma v \rangle\simeq 2\times10^{-33}\rm cm^3 \,s^{-1}$. For a Moore profile ($\gamma=1.5$) this ionization rate is achieved for $\langle \sigma v \rangle\simeq 1.25\times10^{-35}\rm cm^3 \,s^{-1}$ and $m_{\chi} = 10$~MeV. For ease of comparison, we show in Fig.~\ref{fig:Lims} (right axis -- dashed lines) the expected (averaged within the CMZ) ionization rate for a fixed annihilation cross-section of $\langle \sigma v \rangle = 10^{-35}$~cm$^3$/s, with the numerical diffusion calculation (shown also in Fig.~S2 of the Supp. Material).
On top of this, we show that the expected radial profile of the ionization from DM is relatively flat within the CMZ, even for the case of a Moore profile (see more details in Sections II and III of the Supp. Material), as shown in the bottom panel of Fig.~\ref{fig:IonRate}. This is in line with the relatively smooth radial profile of the observed ionization rate.

In terms of constraints from Galactic probes, we remark that X-ray constraints are competitive with those from CMB, although for DM particles with masses below $\sim10$-$20$~MeV, 
diffuse reacceleration of the $e^\pm$ pairs injected by DM~\cite{DelaTorreLuque:2023olp, seo1994stochastic}
induces uncertainties at MeV energies. 
We have tested the X-ray emission produced by the DM-induced $e^{\pm}$ population using the {\it Hermes} code~\cite{Dundovic:2021ryb} (as in Ref.~\cite{DelaTorreLuque:2023olp}), with the same propagation parameters as used for the estimations of the ionization rate in the CMZ, finding that XMM-Newton observations cannot rule out DM with masses below a few tens of MeV, as shown in the right panel of Fig.~S4 (Supp. Material, which contains the additional Refs.~\cite{XMM, Cirelli_2023}). 
We have also calculated the bremsstrahlung emission (using the {\it Hermes} code as in Ref.~\cite{Luque_2024}) at the MeV scale. Although bremsstrahlung emission is expected to be more important for low DM masses~\cite{Cirelli_Bremss} and dominates over the inverse Compton emission for DM masses below a few tens of MeV~\cite{Bartels_2017}, we find that this emission is not expected to be significant for the annihilation strength needed to explain the CMZ ionization rate and it does not represent an important constraint (see right panel of Fig.~S4 in the SM, which contains the additional Ref.~\cite{Bouchet:2010dj}). As expected, the bremsstrahlung emission is not competitive with observables such as the $511$~keV line, X-rays or the CMB. 
In this context, there are indications that point to an excess of diffuse photons at MeV energies~\cite{Karwin_2023}. Annihilation of MeV DM could contribute to the diffuse emission around the GC and help alleviate the current tension with the expected background emission.

\emph{Possible connection with the 511 keV line emission --}
We now investigate the tantalizing possibility that the $511$~keV line measurements around the GC are correlated with the hydrogen ionization rate in the CMZ. Our discussion is motivated by
the following: 

i) Refs.~\cite{Beacom:2005qv, Sizun:2006uh, Sizun:2007ds} have found that the source of the bulge $511$~keV emission must require positrons injected with energies lower than a few MeV, so that only DM with mass below $\lesssim 10$~MeV is able to produce the emission. 

ii) Ref.~\cite{DelaTorreLuque:2023cef} showed that, due to the propagation of positrons, the longitude profile of the $511$~keV line emission can be reproduced only for DM profiles cuspier than the NFW, also favoring low DM masses (see the Appendix of Ref.~\cite{DelaTorreLuque:2023cef} and further discussions in Ref.~\cite{DelaTorreLuque:2024wfz}). Although a contracted NFW distribution would better fit the signal, the exact contraction index (or value of the slope, $\gamma$) depends on the mass of the DM particle annihilating (given that propagation induces a mass-dependent behavior of the 511 keV emission profile from DM) and the exact propagation parameters used.

iii) The power density from DM annihilations in the CMZ is compatible with observations of the line emission at the GC for a cross-section $\langle \sigma v \rangle \lesssim 10^{-32}$~cm$^{3}$/s. For example, a $1$~MeV particle in a Moore DM profile can produce $\sim 10^{52}~\rm erg/s \sim 5\times 10^{50}$$e^+/$yr for $\langle \sigma v \rangle \lesssim 10^{-33}$~cm$^{3}$/s. 

To test whether both can be connected, we compute the $511$~keV emission in the bulge (integrated in the inner $8^\circ$) from the annihilation of sub-GeV DM for the different DM profiles considered in this work, following the same approach as in Ref.~\cite{DelaTorreLuque:2024zsr}, which is consistent with previous evaluations~\cite{Calore:2021klc, Calore:2021lih, DelaTorreLuque:2023huu, DelaTorreLuque:2023nhh, Carenza:2023old}.
Adopting an annihilation cross-section of $\langle \sigma v \rangle = 10^{-35}$~cm$^3$/s, we depict the predicted $511$~keV emission in Fig.~\ref{fig:Lims} (left axis -- solid lines), where we include a shaded gray band that accounts for the observed emission at the bulge~\cite{Siegert:2015knp}, to which we add a conservative $50\%$ uncertainty (to account for underestimated systematic uncertainties). 
From a comparison to the ionization rate at the CMZ (right axis of the figure -- dashed lines), we observe that the annihilation cross-section needed to explain the $511$~line emission at the bulge is lower than the annihilation cross-section that explains the CMZ ionization rate, by at least a factor of a few, meaning that a simultaneous explanation of both phenomena seems difficult under the DM hypothesis. This may indicate that the source of the CMZ ionization rate is not symmetrically injecting electrons and positrons, and could instead be dominated by ionizing protons (or only electrons). 

However, we note here that while uncertainties from propagation parameters similarly affect both predictions, the estimated $511$~keV emission is also affected by the conditions for positronium production. The formation of positronium from thermal positrons is highly dependent on the composition, temperature and density of the medium~\cite{Guessoum1991}, hindering our ability to make a robust prediction of this quantity (especially around the GC). Therefore, it would be difficult to firmly rule out the DM hypothesis of the CMZ ionization rate from the estimated $511$~keV line bulge emission given that uncertainties in this estimate are larger than an order of magnitude and that the cross-sections needed to match both observations can be close (e.g. for the cNFW profile case with 1 MeV DM where the 511 keV emission is compliant at $\phi^{511\,\textrm{keV}}_{8^\circ}\gtrsim 10^{-3}$ cm$^{-2}$ s$^{-1}$ and $\zeta\simeq 10^{-16}\,$s$^{-1}$, only a factor of a few below the observed rate shaded red). 
In addition, we also remark that there are other mechanisms that could boost the ionization rate without affecting the $511$~keV production: invoking more exotic DM scenarios, one can devise a scenario where asymmetric production of electrons and positrons (whilst maintaining charge conservation) can be achieved. 
We also remark that some known astrophysical sources could boost the ionization rate without affecting the $511$ keV production. Another intriguing possibility that may make both anomalies compatible with DM is that the direct interaction of DM particles with H$_2$ molecules also produces ionization.

Since correlations between the $511$~keV line emission and the ionization rate in the CMZ can only happen if the source is injecting electrons or positrons simultaneously, we remark that observations of the ionization rate in the CMZ or even in molecular clouds, could help to discern the nature of the source producing the puzzling $511$~keV flux around the GC. However, it is clear that any source producing the emission of $511$~keV photons at the GC must also play a role in ionizing the CMZ. These correlations will be explored more thoroughly in a separate work. We also note that the broad range of empirical estimates of the molecular ionization rate (potentially favoring a relatively uniform distribution within the CMZ) makes it difficult to compare with
the 511 keV line morphology, which suggests a cusped structure. Moreover, the angular resolution of the SPI detector (and similarly, the proposed COSI~\cite{Siegert:2020oxw} experiment) remains too low (approximately $2.7^{\circ}$) to offer insights into the line's distribution within the inner $\sim200$~pc. Therefore, future observations with improved angular resolution are essential to better understand the emission within the CMZ.

\emph{Discussion and conclusions ---}
In this Letter, we show that the anomalous ionization rate observed at the CMZ can be explained by DM particles of mass below a few tens of MeV annihilating into $e^+e^-$, for DM distributions with slopes slightly above that of the NFW $\gamma>1$. The thermally averaged annihilation 
cross-section required to match the observed ionization rates at the CMZ is below $10^{-33}$~cm$^3$/s for a contracted NFW distribution ($\gamma=1.25$), and can be as low as $10^{-37}$~cm$^3$/s for a Moore DM distribution, all well below the most stringent existing upper limits. 
Using both analytical approximate calculations and a more detailed numerical calculation with the {\tt DRAGON2} code, we show that the required annihilation 
cross-sections needed to explain the CMZ ionization rate are so low that annihilation of sub-GeV DM would produce no detectable inverse Compton, bremsstrahlung or synchrotron emission. Using extreme variations of our diffusion parameters, we find that uncertainties in the predicted ionization rate can be as high as an order of magnitude (see Section S.1.). However, given the low ionization rate required to explain the CMZ ionization rate and the fact that this mass region is largely unconstrained, these uncertainties can be easily compensated by a different value of the annihilation cross-section and provide the required ionization rate in the CMZ even for the most conservative of our estimations.
While the power required to explain the ionization rate in the CMZ is around $10^{41}$~erg/s, DM following a NFW distribution would provide a power of $\sim 10^{44}~\rm erg/s$, in the case of a $1$~MeV particle and $\langle \sigma v \rangle \lesssim 10^{-33}$~cm$^{3}$/s. In a Moore DM profile the power within the CMZ increases above $10^{50}~\rm erg/s$ for the same mass and annihilation cross-section.
Furthermore, one of the strongest points of the DM explanation is that it produces a very flat radial profile of the ionization within the CMZ, as shown in the bottom panel of Fig.~\ref{fig:IonRate}. For a NFW DM distribution, the radial profile of the ionization rate is quite flat, varying by no more than a factor of $3$ for a $5$~MeV DM particle, while much flatter at higher masses, for which the $e^{\pm}$ diffuse faster. For the Moore profile, representing the cuspiest DM distribution considered here, the ionization rate at different radial distances varies between a factor of a few and slightly more than an order of magnitude, values which are still compatible with the observations of the CMZ.

The fact that sub-GeV DM can even exceed the observed ionization rate in the CMZ without being in conflict with these probes demonstrates that this observable can be a very powerful tool for constraining sub-GeV DM. 
We find that the annihilation cross-section needed to explain the CMZ ionization rate is above, by at least a factor of a few, that needed to explain the bulge $511$~keV emission, whilst emphasizing that the uncertainties in the 511 keV emission make it difficult to presently rule out DM as the source of ionization in the CMZ. Therefore, we determine that the $511$~keV emission cannot conclusively rule out the DM explanation of the CMZ ionization rate for DM masses below a few MeV, leaving open a window to explore whether both observations are related (even beyond the DM hypothesis).

We conclude that the annihilation of sub-GeV DM that can explain the ionization rate in the CMZ could significantly contribute to an outstanding astrophysical mystery, namely the nature of the $511$~keV line emission. Finding spatial correlations between molecular clouds and positron line emission, as well as X-ray, $\gamma$ ray and radio constraints on correlated emissions may enable us to ascertain if the source of molecular ionization must be leptonic.

\emph{Acknowledgements---} 
We would like to thank Marco Cirelli for helpful feedback on the draft of this letter. SB is supported by the STFC under grant ST/X000753/1. PDL is supported by the Juan de la Cierva JDC2022-048916-I grant, funded by MCIU/AEI/10.13039/501100011033 European Union "NextGenerationEU"/PRTR. The work of PDL is also supported by the grants PID2021-125331NB-I00 and CEX2020-001007-S, both funded by MCIN/AEI/10.13039/501100011033 and by ``ERDF A way of making Europe''. PDL also acknowledges the MultiDark Network, ref. RED2022-134411-T. This project used computing resources from the Swedish National Infrastructure for Computing (SNIC) under project No.2022/3-27 partially funded by the Swedish Research Council through grant no. 2018-05973.

\newpage

\appendix
\onecolumngrid

\section{Propagation setup and calculations}
\label{sec:Setups}
As discussed in the main text, we used two different approaches to compute the steady-state diffuse distribution of electrons and positrons injected by DM in the CMZ. While the analytical approximation allows us easily estimate the average ionization within the CMZ, we need to use a numerical scheme to account for the effects of diffusion in the $e^{\pm}$. The details are given below.

\emph{Analytical approximation ---}
In this approximation, we neglect diffusion of the positrons and electrons within the CMZ and calculate their steady-state distribution as 
\begin{equation}
    \frac{d\phi_e}{dE}\left(E, \textbf{x} \right) = \frac{1}{b_e(E, n(\textbf{x}))}\int Q_e (E', \textbf{x}) dE' \, ,
    \label{eq:Anal}
\end{equation}
with $b_e$ as the energy loss term ($b_e \equiv \frac{dE}{dt}^\textrm{ion}$), calculated as implemented in the {\tt DRAGON2} code (see Eq.~(C.39) of Ref.~\cite{DRAGON2-1}), $n(x_\textrm{CMZ})=150$~cm$^{-3}$ is taken as the mean density of the CMZ~\cite{IoRCR} and $Q_e\left(E, \textbf{x} \right)$ as the injection spectrum (source term) from DM annihilation, which takes the following form \cite{Cirelli:2010xx}
\begin{equation}
     Q_e\left(E, \textbf{x} \right) = \frac{\langle \sigma v\rangle}{2} \left(\frac{\rho_\chi(\textbf{x})}{m_\chi}\right)^2\frac{dN_e^{\textrm{ann}}}{dE_e} \, ,
\label{eq:Source}    
\end{equation}
where $\langle \sigma v\rangle$ is the DM thermally averaged annihilation cross-section, $m_{\chi}$ is the DM mass, $\rho_{\chi}$ is the DM density in the CMZ and $\frac{dN_e^\textrm{ann}}{d E_e }= \delta(E + m_e - m_{\chi})$.

We notice that the approximation used to obtain Eq.~\eqref{eq:Anal} is only valid for low energy particles below a few tens of MeV and in very dense media, but the simplifying advantage of this estimation is that it does not depend on the diffusion coefficient or other propagation parameters that are not precisely known near the GC.

\emph{Numerical calculation ---} 
In our main approach, we compute the steady-state diffuse $e^{\pm}$ spatial distribution and energy spectra with a recent version~\cite{de_la_torre_luque_2023_10076728} of the {\tt DRAGON2} code~\cite{DRAGON2-1, DRAGON2-2}, a dedicated CR propagation code designed to numerically solve the full diffusion-advection-reacceleration-loss equation for the transport of charged particles in the Galactic environment~\cite{Ginz&Syr}. Here, we incorporate two main changes to the {\tt DRAGON2} code: the implementation of a logarithmic spatial grid that allows us to reach sub-pc spatial scales as well as the inclusion, in the neutral gas distribution, of a region resembling the CMZ, with density set to be constant at 150 cm$^3$ and spanning the size of the CMZ. 
We use the same prescription and propagation parameters as used in Refs.~\cite{DelaTorreLuque:2023huu, DelaTorreLuque:2023nhh, DelaTorreLuque:2023olp, DelaTorreLuque:2023cef}, where we model the fluxes of sub-GeV positrons from DM and other exotic positron sources.

We compute the steady-state diffuse distribution of electrons and positrons injected by DM by solving the diffusion-advection-convection-loss equation~\cite{Ginz&Syr, DRAGON2-1}
\begin{widetext}
    \begin{equation}
    \label{eq:CRtransport}
        - \nabla\cdot\left(D\vec{\nabla} f_e + \vec{v}_c f_e \right) - \frac{\partial}{\partial p_e} \left[\dot{p}_e f_e - p_e^2 D_{pp} \frac{\partial}{\partial p_e}\left(\frac{f_e}{p_e^2}\right) - \frac{p_e}{3}\left(\vec{\nabla}\cdot\vec{v}_c f_e\right)\right] = Q_e\;,  
    \end{equation}
\end{widetext}
\onecolumngrid
which is solved, using a Crank–Nicolson scheme~\cite{Crank_Nicolson_1947}, for $f_e \equiv \frac{dn_e}{dp_e}$, the density of $e^\pm$ per unit momentum at a given position. This equation includes: i) spatial diffusion with diffusion coefficient $D$, ii) momentum losses $\dot{p}_i$ due to interactions with the Galactic environment, iii) momentum diffusion (or reacceleration) with diffusion coefficient $D_{pp}$, iv) convection due to Galacitc winds $\vec{v}_c$. 
Here, we note that the DM density in the GC is cut at $r < 2 R_\textrm{s}$, where $R_\textrm{s}$ is the Schwarzschild radius of Sgr A*, the supermassive black hole in the GC~\cite{EventHorizonTelescope:2022wkp}. This is an important physical boundary condition to prevent the source term (Eq.~\eqref{eq:Source}) diverging at $r=0$.
Here, we adopt a spatial grid with a bin size varying from $0.1$~pc in the inner Galaxy to $\simeq 200$~pc, increasing progressively as the radial distance from the GC is increased. 
 We simulate electron-positron signals from DM annihilation over a range of kinetic energies from $1$~keV to a few GeV, with an energy resolution of 5\%. We ensure convergence of the simulations using a variable time step (see Sect.~4.5 of Ref.~\cite{DRAGON2-1}), with a minimum time step of $0.1$~kyr and a maximum time step of $64$~Gyr, that cover the relevant timescales involved, similar to what was done in Ref.~\cite{DelaTorreLuque:2023cef}.

The diffusion coefficient employed is determined  through a combined fit to CR measurements by AMS-02~\cite{AGUILAR20211, aguilar2018observation, AMS_gen} and Voyager-1~\cite{VoyagerMO, Stone150}, and is given by  $D (R) = D_0 \beta^{\eta}\left(R/R_0 \right)^{\delta}$, where $D_0 = 1.2\times10^{29}$~cm$^2$/s is its normalization at $R_0 = 4$~GV, $\beta$ is the speed of the particles in units of the speed of light, $\eta = -0.75$ and $\delta = 0.49$. These are found for a halo height of $8$~kpc. We use these as our reference propagation parameters but also compare the results obtained with extreme variations in the next section (Sec.~\ref{sec:AppTimeScales}). Unfortunately, we have no information about the exact propagation parameters within the CMZ, however, the use of extreme variations of the parameters obtained from CR analyses serves to estimate how much our results can change from our reference setup.
We refer to Ref.~\cite{DelaTorreLuque:2023olp} for full details on the impact of the propagation parameters in the flux of the electrons and positrons injected by sub-GeV DM.

The advantage of this approach is that it is able to account for the effects of diffusion and reacceleration, but there are two main caveats: the first is that it depends on the extrapolation of the propagation parameters obtained from analyses of local CR data at GeV energies (see Ref.~\cite{DelaTorreLuque:2023olp} for detailed discussion), and the second is that the code does not consider how the dynamics of diffusion can change when crossing the border of the CMZ. Regarding the latter, Ref.~\cite{Phan_2023} showed that crossing a very dense medium can lead to a reduction of the flux of particles within molecular clouds, thereby reducing the ionization rate. 
This effect is more pronounced for lower positron energies and hence lower DM masses since they produce lower energy positrons. However, we note that this effect is not so relevant in our case, given that the electrons and positrons producing ionization come mostly from the interior of the CMZ, and that even with high suppression of the $e^{\pm}$ flux the predicted ionization rate can exceed that observed at the CMZ by orders of magnitude, which will be discussed further below. 
Therefore, differences in the predicted ionization rates from both approaches are expected, especially at masses above a few tens of MeV where the analytical approximation is not valid anymore because the diffusion rate starts to become dominant (see Fig.~\ref{fig:TimeScales} in the next section).

\section{Relevant timescales and propagation uncertainties}
\label{sec:AppTimeScales}

\begin{figure*}[h!]
\centering 
\includegraphics[width=0.5\linewidth]{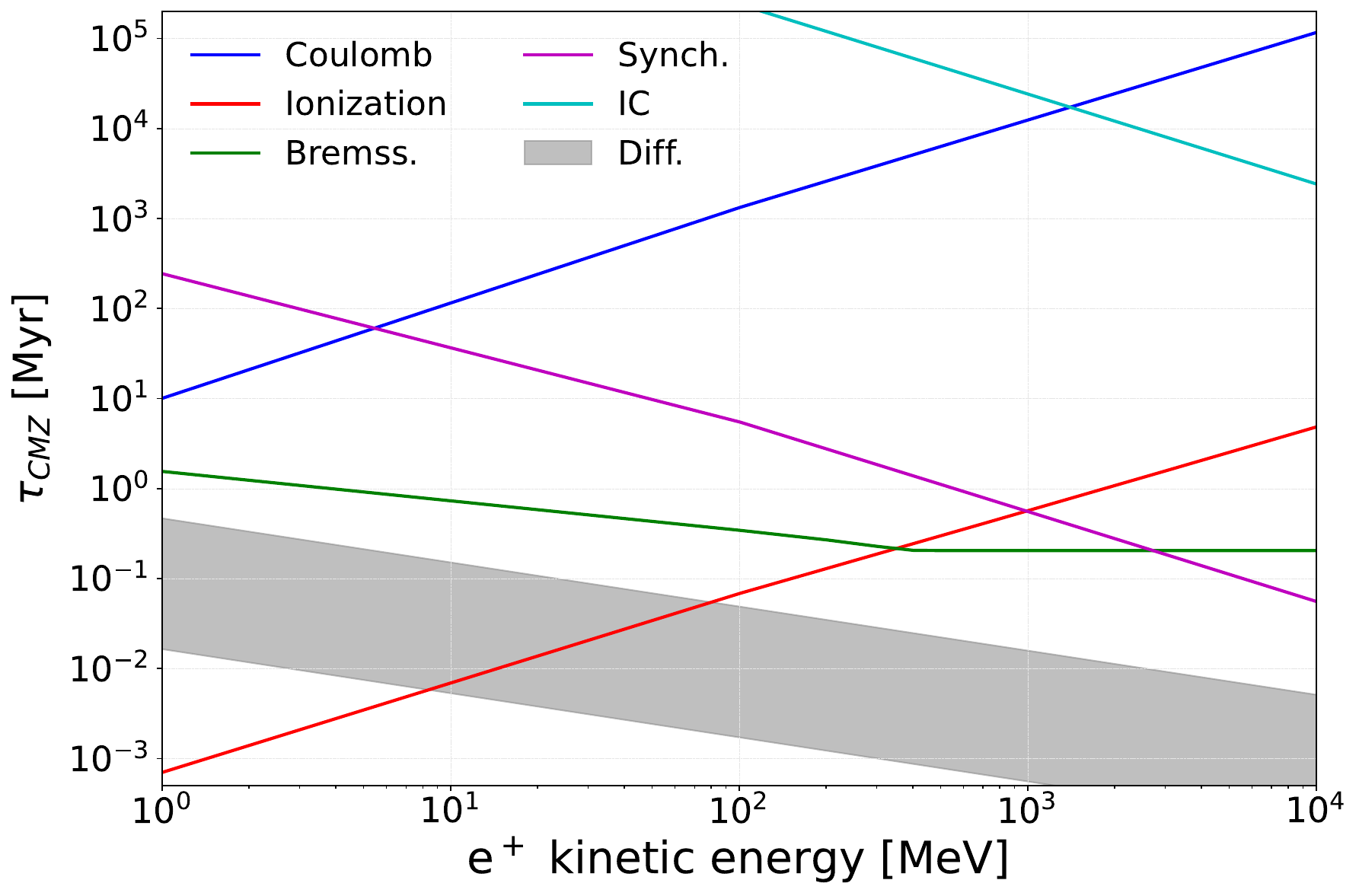}
\caption{Main timescales involved in the propagation of electrons and positrons generated by sub-GeV DM particles, for a gas density of $150$~cm$^{-3}$ composed of H$_2$ and  $10\%$ of He. The magnetic field strength here is set to $150$~$\mu$G and the energy density of the interstellar radiation fields is taken from~\cite{Evoli_2020}. The grey band indicates the difference in diffusion time in the Optimistic and Pessimistic setups, used to calculate uncertainties from diffusion in our estimations of the CMZ ionization rate.}
\label{fig:TimeScales}
\end{figure*}

Our calculations include all the relevant Galactic CR propagation processes in a broad energy range. For signals from sub-GeV DM, it is crucial to account for diffusion, reacceleration, and energy losses, that become dominant below a few tens of MeV. The most relevant cooling processes at these energies are Coulomb and ionization losses, although we include synchrotron, bremsstrahlung, inverse Compton and catastrophic losses
(annihilation) in our simulation as well.
A comparison of the relevant timescales is illustrated in Fig.~\ref{fig:TimeScales}, where we show a band around the benchmark diffusion time (i.e. diffusion time of particles with our benchmark diffusion setup) to indicate the difference between two extreme diffusion setups, the optimistic and pessimistic scenarios, which are explained below. Moreover, cooling is much faster than diffusion. We note that the average distance traveled a positron injected at $10$~MeV (which can be calculated as $\langle l_{\text{travel}} \rangle \approx \sqrt{2D\tau}$, being $D$ the diffusion coefficient and $\tau$ the dominant timescale) is of tens of parsecs in a $150$~cm$^{-3}$ gas, without accounting for wind advection or reacceleration (that can boost $\langle l_{\text{travel}} \rangle$ to even more than hundred parsecs).

Given that the propagation conditions around the GC are currently so challenging to evaluate, we employ the propagation parameters that reproduce the local CR observations, in particular AMS-02 and Voyager-1 data, as explained in detail in Ref.~\cite{DelaTorreLuque:2023olp}. This allows us to make a reasonable estimate of the expected ionization rate in the CMZ. Similarly to what was done in Ref.~\cite{DelaTorreLuque:2023olp}, in order to understand the impact of uncertainties in the propagation parameters, we have studied two opposite and extremal setups:
\begin{itemize}
    \item The ``optimistic'' setup, where CRs diffuse much faster than in our reference setup. Here, the normalization of the diffusion coefficient, $D_0$, is doubled (i.e. in this setup $D_0 = 20.4 \times 10^{28}$~cm/s$^2$) with respect to our reference value and have high reacceleration (which effectively contributes to a faster diffusion since it promotes low energy particles to higher energy particles), with $V_A = 40$~km/s.
    
    \item The ``pessimistic'' scenario, where diffusion is significantly reduced - to a value of $3.8\times 10^{28}$~cm/s$^2$ - and there is no reacceleration.
\end{itemize}

\begin{figure*}[t!]
\centering 
\includegraphics[width=0.48\linewidth]{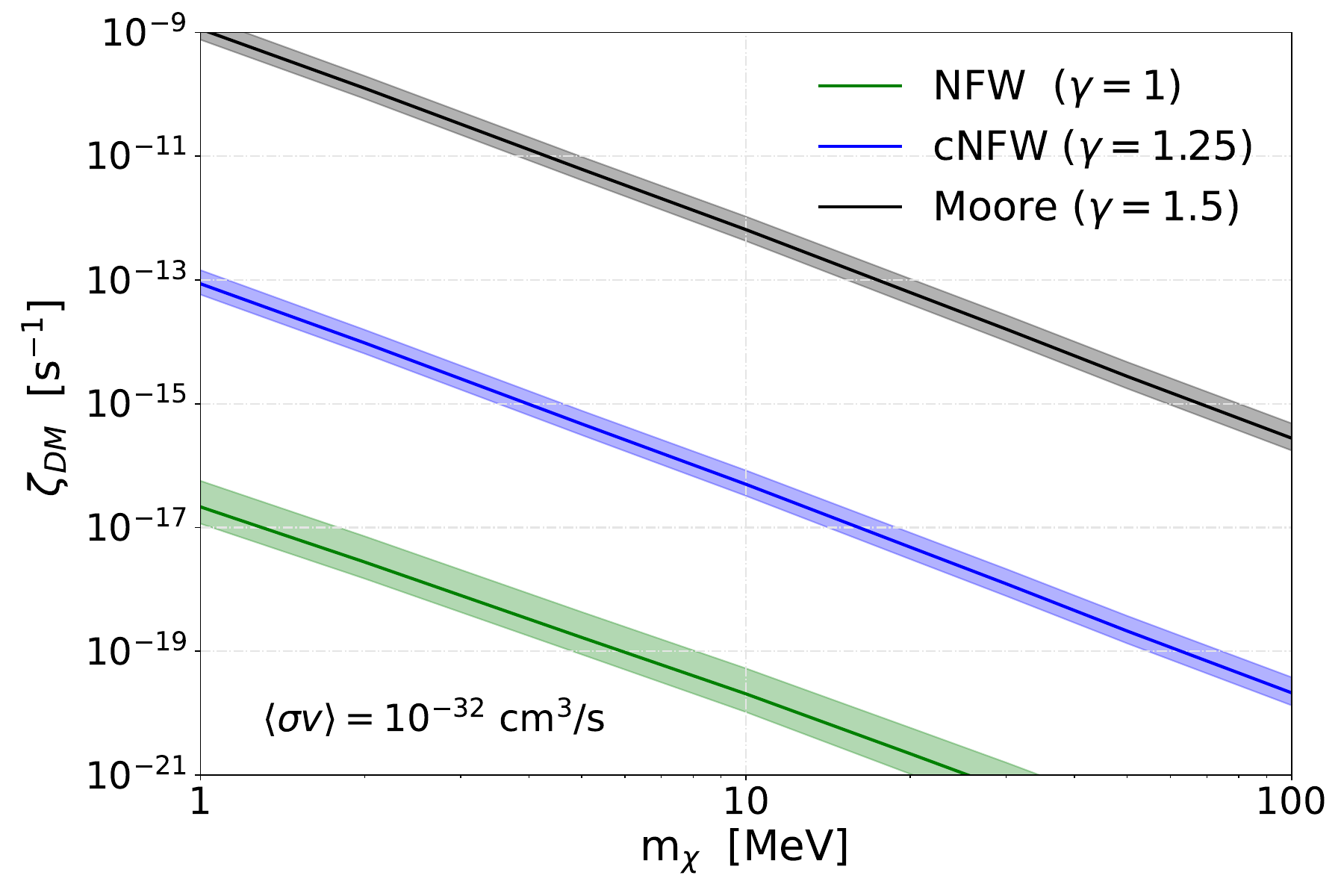}
\includegraphics[width=0.48\linewidth]{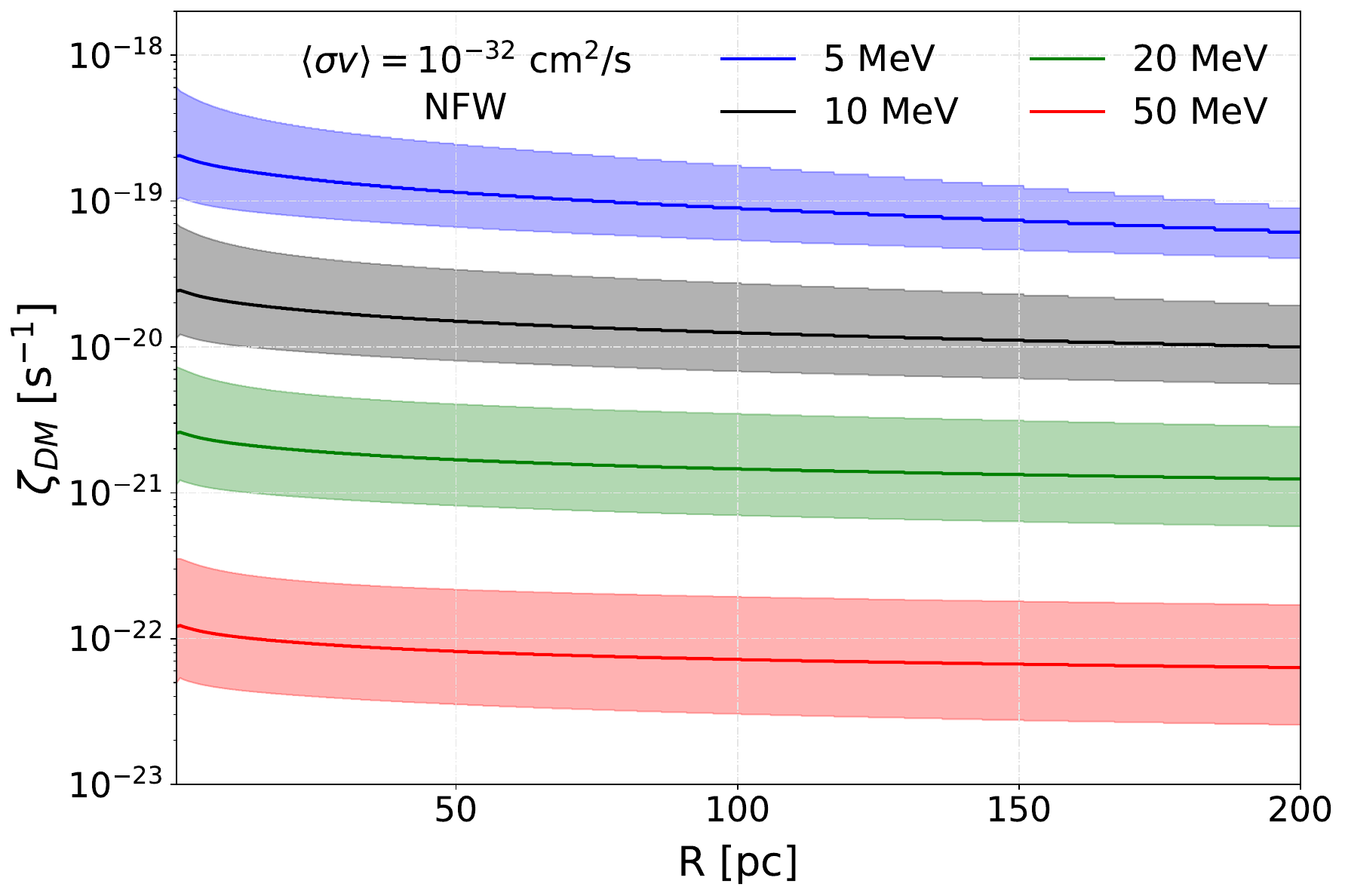}
\caption{\textbf{Left panel:} Uncertainties in the average ionization rate from DM for different DM distributions. The bands represent the different from the estimation with the Optimistic and Pessimistic setups. 
\textbf{Right panel:} Similar to the left panel but for the radial profile of the ionization, and for an NFW profile.}
\label{fig:PropUnc}
\end{figure*}

As we show in the left panel of Fig.~\ref{fig:PropUnc}, the average ionization rate predicted in the CMZ changes by a factor of a few in our estimations, being more important at higher masses than at lower masses. This is because, in the MeV energy region, the dominant timescale for propagation is energy losses, which reduces the impact of uncertainties in diffusion. For completeness, we show, in the left panel of Fig.~\ref{fig:PropUnc}, uncertainties in the expected radial profile of the ionization rate (assuming an NFW DM distribution), where this becomes clearer. 
These uncertainties can be easily compensated for with a different value of the annihilation cross-section and provide the required ionization rate in the CMZ even for the most conservative of our estimations, since the annihilation cross-section required is largely unconstrained in this mass range.

\section{Ionization rate for fixed cross-sections}
\label{sec:IoRFixed}

\begin{figure}[h!]
\includegraphics[width=0.49\linewidth]{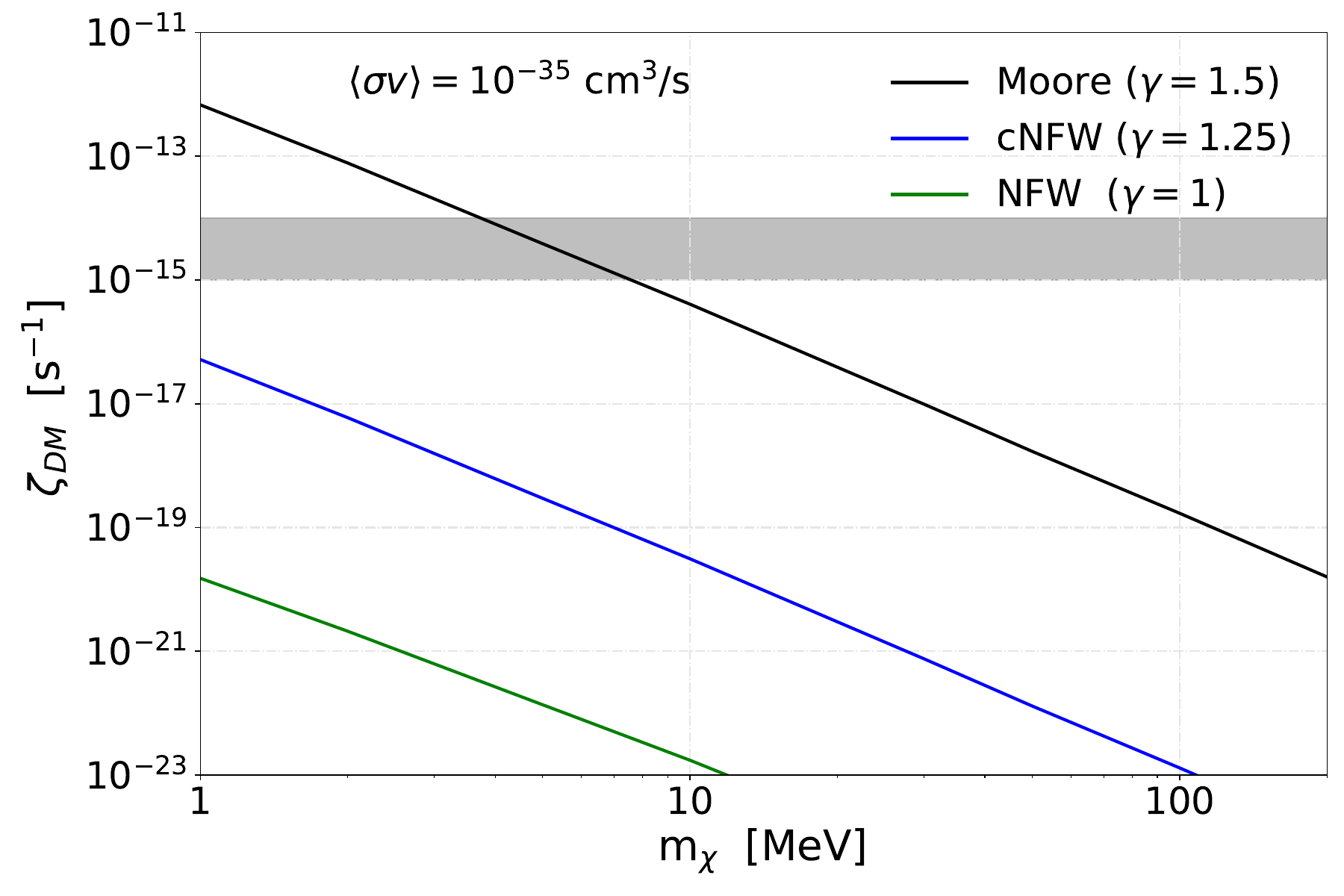}

\caption{Expected ionization rate in the CMZ from annihilation of sub-GeV DM particles into $e^+e^-$, as obtained with the numerical calculation and for different DM distributions. The band illustrates the range of observed values of ionization rate in the CMZ. In particular, we show the average ionization rate within the CMZ for $\langle \sigma v \rangle = 10^{-35}$~cm$^3$/s. The different colors refer to the DM distributions indicated in the legend. }
\label{fig:FixedIoR}
\end{figure}

For ease of comparison, we show in  Fig.~\ref{fig:FixedIoR} the predicted ionization rate in the CMZ from sub-GeV DM annihilation, for different DM distributions and a fixed annihilation cross-section of $\langle \sigma v \rangle = 10^{-35}$~cm$^3$/s. 
As shown in the bottom panel of Fig.~1, the predicted ionization profiles become very flat when including the effects of propagation. In this figure, we show the profile of the ionization rate for different DM masses, comparing the predictions using a Moore profile and an NFW one. From this figure, one can see that for a NFW DM distribution, the radial profile of the ionization rate is quite flat, varying by no more than a factor of 3 for a 5 MeV DM particle, while much flatter at higher masses, for which the $e^{\pm}$ diffuse faster. In comparison, the ionization rate in the CMZ, measured employing different techniques, has been observed to be scattered within values that cover two orders of magnitude~\cite{IoRCR}. For the Moore profile, representing the cuspiest DM distribution used here, the ionization rate at different radial distances varies between a factor of a few and slightly more than an order of magnitude, which can be still compatible with the observations. Therefore, we conclude that DM is expected to produce smooth ionization profiles in the CMZ as measured. Point-like sources or Sgr A* fail on this point. However, a smooth population of sources injecting low-energy particles within the CMZ can provide a similarly good explanation.

\section{X and gamma ray emissions from annihilation of sub-GeV DM}
\label{sec:XGamma}

\begin{figure*}[h!]
\includegraphics[width=0.49\linewidth, height=0.24\textheight]{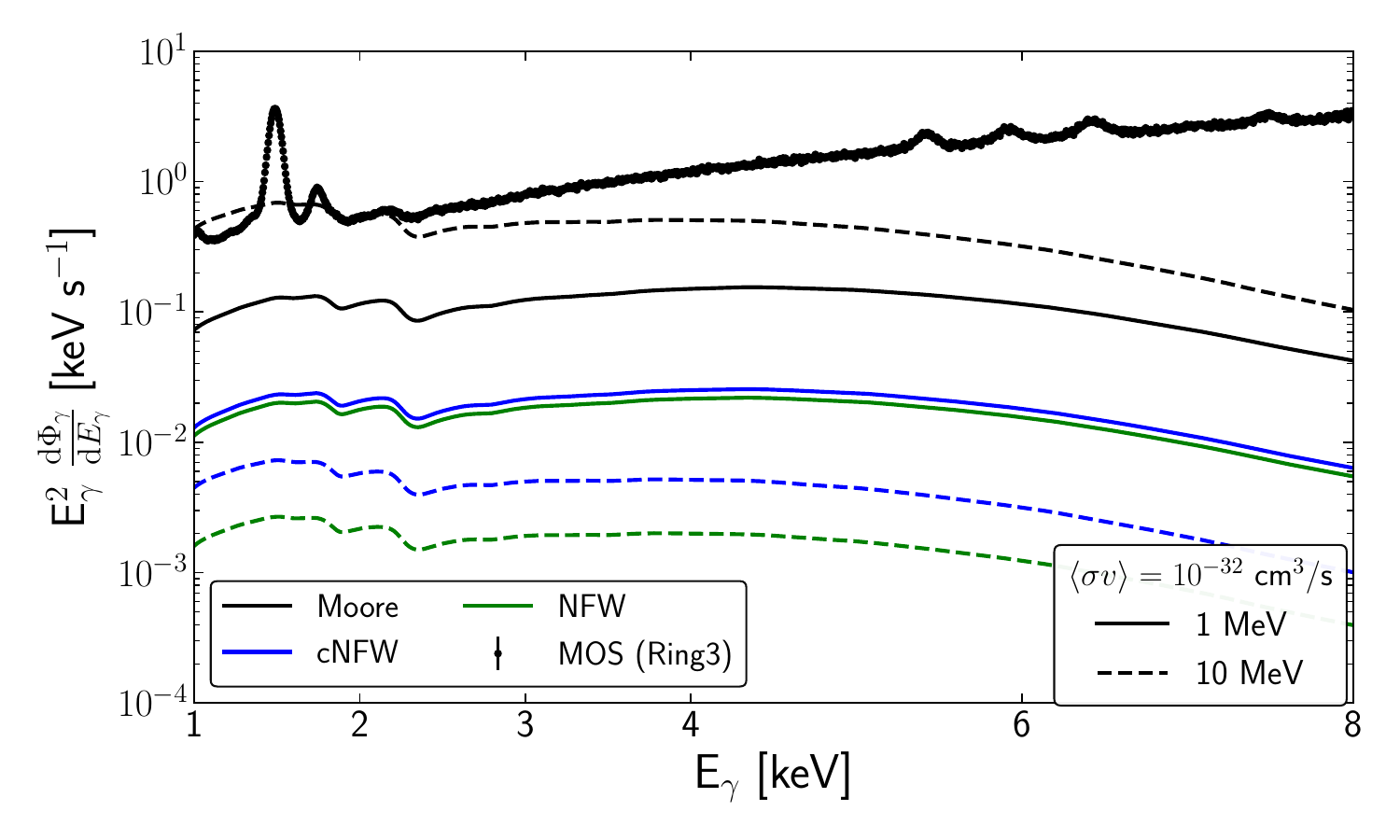}
\includegraphics[width=0.49\linewidth, height=0.24\textheight]{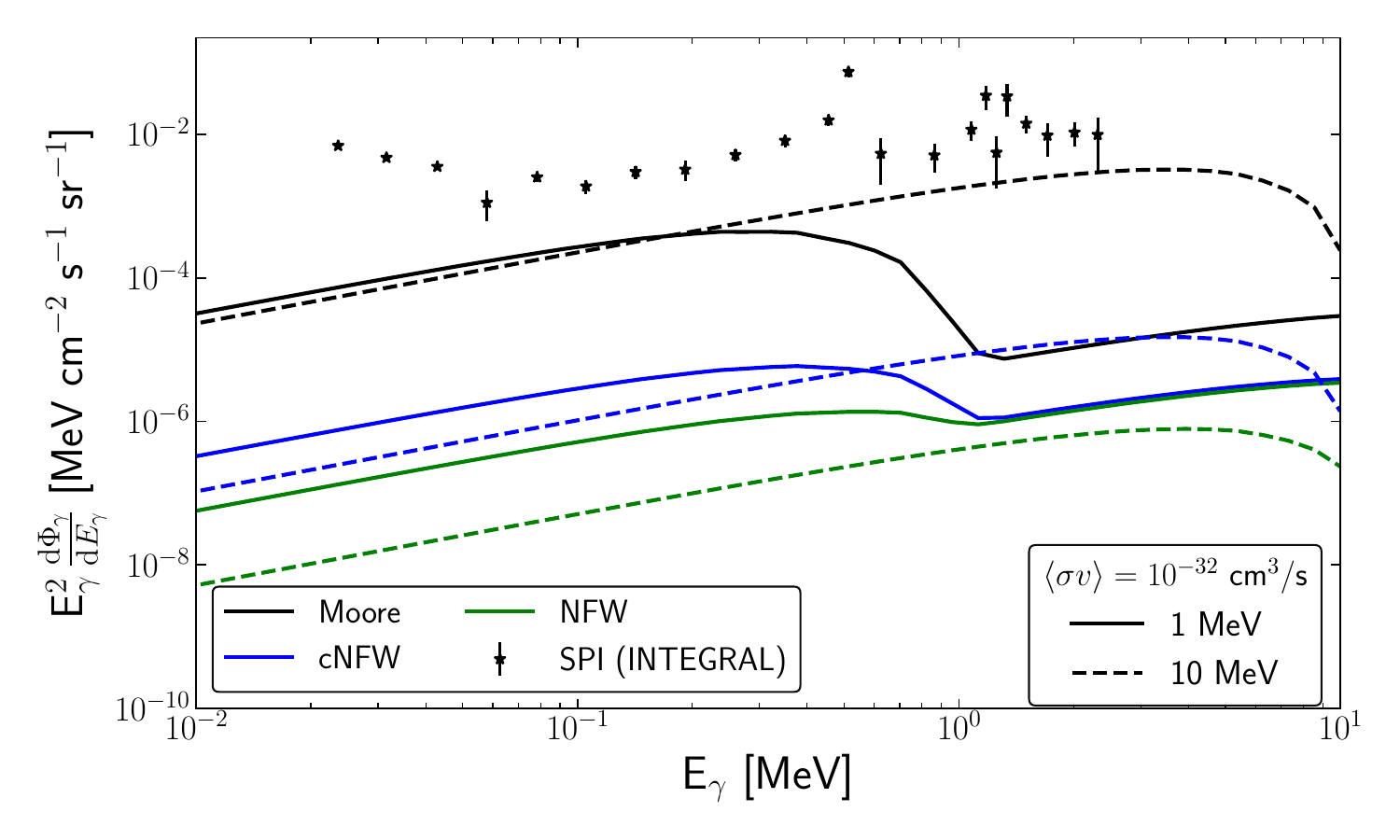}
\caption{\textbf{Left panel:} X-ray emission associated to the $e^{\pm}$ population from DM compared to XMM-Newton data in the most constraining ring (Ring 3)~\cite{XMM}. \textbf{Right panel:} Bremsstrahlung emission associated with the $e^{\pm}$ population injected by DM compared to the diffuse MeV $\gamma$-ray data from SPI (INTEGRAL)~\cite{Bouchet:2010dj, Siegert:2015knp} in the $|b|<15^{\circ}$, $|l|<30^{\circ}$ region. The $\gamma$-ray flux at energies above the DM mass is due to reacceleration, that boosts the $e^{\pm}$ population to higher energies~\cite{DelaTorreLuque:2023olp}. In both cases, these estimations are made for a $1$ and $10$~MeV DM particle for $\langle \sigma v \rangle = 10^{-32}$~cm$^3$/s, for different DM profiles}
\label{fig:XBremss}
\end{figure*}

In the left panel of Fig.~\ref{fig:XBremss}, we compare the predicted X-ray emission (which comes mostly from inverse Compton scattering of the DM-induced $e^{\pm}$ population with CMB and other interstellar radiation fields) from the annihilation of $1$ and $10$~MeV DM for different profiles and for $\langle \sigma v \rangle = 10^{-35}$~cm$^3$/s, with the most constraining data set of the XMM-Newton measurements derived in Ref.~\cite{XMM}, the Ring 3~\cite{DelaTorreLuque:2023olp, DelaTorreLuque:2023nhh}. The emission from the $e^{\pm}$ particles injected by $10$~MeV DM peak at a few keV, while the peak emission for the $1$~MeV DM products peaks below the keV scale~\cite{Cirelli_2023, DelaTorreLuque:2023olp}. Since the infrared, optical and UV radiation fields are more concentrated around the GC, the inverse-Compton emission from the scattering with these fields is more intense for the cuspier DM profiles.
As a result, annihilation of sub-GeV DM can explain the observed ionization rate at the CMZ without being in conflict with X-ray emissions.
We remark here that the strongest constraints so far on the annihilation rate of sub-GeV DM come from XMM-Newton measurements~\cite{Cirelli_2023, DelaTorreLuque:2023olp} of the X-ray Galactic diffuse emission, impact of the $e^{\pm}$ injection from DM annihilation on the CMB anisotropies~\cite{Slatyer_2016, Lopez_Honorez_2013}, and the longitude profile of the $511$~keV line~\cite{DelaTorreLuque:2023cef}. However, there are systematic uncertainties in the derivation of these limits that make them be less robust than the CMB limits. 

Similarly, in the right panel of this figure we show the predicted MeV emission, dominated by bremsstrahlung, which is calculated with the {\it Hermes} code~\cite{Dundovic:2021ryb}, and compare it to SPI (INTEGRAL) measurements of the diffuse $\gamma$-ray flux. We warn the reader that here we are comparing with the intensity averaged over a large region of the Galaxy. Measurements focused on the CMZ are expected to lead to a significantly higher intensity. According to our results, this does not appear as an issue for the DM hypothesis.

\newpage

\bibliographystyle{apsrev4-1}
\bibliography{references.bib}

\begin{thebibliography}{83}%
\makeatletter
\providecommand \@ifxundefined [1]{%
 \@ifx{#1\undefined}
}%
\providecommand \@ifnum [1]{%
 \ifnum #1\expandafter \@firstoftwo
 \else \expandafter \@secondoftwo
 \fi
}%
\providecommand \@ifx [1]{%
 \ifx #1\expandafter \@firstoftwo
 \else \expandafter \@secondoftwo
 \fi
}%
\providecommand \natexlab [1]{#1}%
\providecommand \enquote  [1]{``#1''}%
\providecommand \bibnamefont  [1]{#1}%
\providecommand \bibfnamefont [1]{#1}%
\providecommand \citenamefont [1]{#1}%
\providecommand \href@noop [0]{\@secondoftwo}%
\providecommand \href [0]{\begingroup \@sanitize@url \@href}%
\providecommand \@href[1]{\@@startlink{#1}\@@href}%
\providecommand \@@href[1]{\endgroup#1\@@endlink}%
\providecommand \@sanitize@url [0]{\catcode `\\12\catcode `\$12\catcode
  `\&12\catcode `\#12\catcode `\^12\catcode `\_12\catcode `\%12\relax}%
\providecommand \@@startlink[1]{}%
\providecommand \@@endlink[0]{}%
\providecommand \url  [0]{\begingroup\@sanitize@url \@url }%
\providecommand \@url [1]{\endgroup\@href {#1}{\urlprefix }}%
\providecommand \urlprefix  [0]{URL }%
\providecommand \Eprint [0]{\href }%
\providecommand \doibase [0]{http://dx.doi.org/}%
\providecommand \selectlanguage [0]{\@gobble}%
\providecommand \bibinfo  [0]{\@secondoftwo}%
\providecommand \bibfield  [0]{\@secondoftwo}%
\providecommand \translation [1]{[#1]}%
\providecommand \BibitemOpen [0]{}%
\providecommand \bibitemStop [0]{}%
\providecommand \bibitemNoStop [0]{.\EOS\space}%
\providecommand \EOS [0]{\spacefactor3000\relax}%
\providecommand \BibitemShut  [1]{\csname bibitem#1\endcsname}%
\let\auto@bib@innerbib\@empty
\bibitem [{\citenamefont {Lopez-Honorez}\ \emph {et~al.}(2013)\citenamefont
  {Lopez-Honorez}, \citenamefont {Mena}, \citenamefont {Palomares-Ruiz},\ and\
  \citenamefont {Vincent}}]{Lopez_Honorez_2013}%
  \BibitemOpen
  \bibfield  {author} {\bibinfo {author} {\bibfnamefont {L.}~\bibnamefont
  {Lopez-Honorez}}, \bibinfo {author} {\bibfnamefont {O.}~\bibnamefont {Mena}},
  \bibinfo {author} {\bibfnamefont {S.}~\bibnamefont {Palomares-Ruiz}}, \ and\
  \bibinfo {author} {\bibfnamefont {A.~C.}\ \bibnamefont {Vincent}},\ }\href
  {\doibase 10.1088/1475-7516/2013/07/046} {\bibfield  {journal} {\bibinfo
  {journal} {Journal of Cosmology and Astroparticle Physics}\ }\textbf
  {\bibinfo {volume} {2013}},\ \bibinfo {pages} {046–046} (\bibinfo {year}
  {2013})}\BibitemShut {NoStop}%
\bibitem [{\citenamefont {Slatyer}(2016)}]{Slatyer_2016}%
  \BibitemOpen
  \bibfield  {author} {\bibinfo {author} {\bibfnamefont {T.~R.}\ \bibnamefont
  {Slatyer}},\ }\href {\doibase 10.1103/PhysRevD.93.023527} {\bibfield
  {journal} {\bibinfo  {journal} {Phys. Rev. D}\ }\textbf {\bibinfo {volume}
  {93}},\ \bibinfo {pages} {023527} (\bibinfo {year} {2016})}\BibitemShut
  {NoStop}%
\bibitem [{\citenamefont {Evoli}\ \emph {et~al.}(2017)\citenamefont {Evoli},
  \citenamefont {Gaggero}, \citenamefont {Vittino}, \citenamefont
  {Di~Bernardo}, \citenamefont {Di~Mauro}, \citenamefont {Ligorini},
  \citenamefont {Ullio},\ and\ \citenamefont {Grasso}}]{DRAGON2-1}%
  \BibitemOpen
  \bibfield  {author} {\bibinfo {author} {\bibfnamefont {C.}~\bibnamefont
  {Evoli}}, \bibinfo {author} {\bibfnamefont {D.}~\bibnamefont {Gaggero}},
  \bibinfo {author} {\bibfnamefont {A.}~\bibnamefont {Vittino}}, \bibinfo
  {author} {\bibfnamefont {G.}~\bibnamefont {Di~Bernardo}}, \bibinfo {author}
  {\bibfnamefont {M.}~\bibnamefont {Di~Mauro}}, \bibinfo {author}
  {\bibfnamefont {A.}~\bibnamefont {Ligorini}}, \bibinfo {author}
  {\bibfnamefont {P.}~\bibnamefont {Ullio}}, \ and\ \bibinfo {author}
  {\bibfnamefont {D.}~\bibnamefont {Grasso}},\ }\href {\doibase
  10.1088/1475-7516/2017/02/015} {\bibfield  {journal} {\bibinfo  {journal}
  {JCAP}\ }\textbf {\bibinfo {volume} {02}},\ \bibinfo {pages} {015} (\bibinfo
  {year} {2017})},\ \Eprint {http://arxiv.org/abs/1607.07886} {arXiv:1607.07886
  [astro-ph.HE]} \BibitemShut {NoStop}%
\bibitem [{\citenamefont {Evoli}\ \emph {et~al.}(2018)\citenamefont {Evoli},
  \citenamefont {Gaggero}, \citenamefont {Vittino}, \citenamefont {Di~Mauro},
  \citenamefont {Grasso},\ and\ \citenamefont {Mazziotta}}]{DRAGON2-2}%
  \BibitemOpen
  \bibfield  {author} {\bibinfo {author} {\bibfnamefont {C.}~\bibnamefont
  {Evoli}}, \bibinfo {author} {\bibfnamefont {D.}~\bibnamefont {Gaggero}},
  \bibinfo {author} {\bibfnamefont {A.}~\bibnamefont {Vittino}}, \bibinfo
  {author} {\bibfnamefont {M.}~\bibnamefont {Di~Mauro}}, \bibinfo {author}
  {\bibfnamefont {D.}~\bibnamefont {Grasso}}, \ and\ \bibinfo {author}
  {\bibfnamefont {M.~N.}\ \bibnamefont {Mazziotta}},\ }\href {\doibase
  10.1088/1475-7516/2018/07/006} {\bibfield  {journal} {\bibinfo  {journal}
  {JCAP}\ }\textbf {\bibinfo {volume} {07}},\ \bibinfo {pages} {006} (\bibinfo
  {year} {2018})},\ \Eprint {http://arxiv.org/abs/1711.09616} {arXiv:1711.09616
  [astro-ph.HE]} \BibitemShut {NoStop}%
\bibitem [{\citenamefont {De~La
  Torre~Luque}(2023)}]{de_la_torre_luque_2023_10076728}%
  \BibitemOpen
  \bibfield  {author} {\bibinfo {author} {\bibfnamefont {P.}~\bibnamefont
  {De~La Torre~Luque}},\ }\href {\doibase 10.5281/zenodo.10076728} {\enquote
  {\bibinfo {title} {Dragon2\_optimized\_dm\&antinuclei},}\ } (\bibinfo {year}
  {2023})\BibitemShut {NoStop}%
\bibitem [{\citenamefont {{Ravikularaman}}\ \emph {et~al.}(2024)\citenamefont
  {{Ravikularaman}}, \citenamefont {{Recchia}}, \citenamefont {{Phan}},\ and\
  \citenamefont {{Gabici}}}]{IoRCR}%
  \BibitemOpen
  \bibfield  {author} {\bibinfo {author} {\bibfnamefont {S.}~\bibnamefont
  {{Ravikularaman}}}, \bibinfo {author} {\bibfnamefont {S.}~\bibnamefont
  {{Recchia}}}, \bibinfo {author} {\bibfnamefont {V.~H.~M.}\ \bibnamefont
  {{Phan}}}, \ and\ \bibinfo {author} {\bibfnamefont {S.}~\bibnamefont
  {{Gabici}}},\ }\href {\doibase 10.48550/arXiv.2406.15260} {\bibfield
  {journal} {\bibinfo  {journal} {arXiv e-prints}\ ,\ \bibinfo {eid}
  {arXiv:2406.15260}} (\bibinfo {year} {2024})},\ \Eprint
  {http://arxiv.org/abs/2406.15260} {arXiv:2406.15260 [astro-ph.HE]}
  \BibitemShut {NoStop}%
\bibitem [{\citenamefont {Miller}\ \emph {et~al.}(2020)\citenamefont {Miller},
  \citenamefont {Tennyson}, \citenamefont {Geballe},\ and\ \citenamefont
  {Stallard}}]{RevModPhys.92.035003}%
  \BibitemOpen
  \bibfield  {author} {\bibinfo {author} {\bibfnamefont {S.}~\bibnamefont
  {Miller}}, \bibinfo {author} {\bibfnamefont {J.}~\bibnamefont {Tennyson}},
  \bibinfo {author} {\bibfnamefont {T.~R.}\ \bibnamefont {Geballe}}, \ and\
  \bibinfo {author} {\bibfnamefont {T.}~\bibnamefont {Stallard}},\ }\href
  {\doibase 10.1103/RevModPhys.92.035003} {\bibfield  {journal} {\bibinfo
  {journal} {Rev. Mod. Phys.}\ }\textbf {\bibinfo {volume} {92}},\ \bibinfo
  {pages} {035003} (\bibinfo {year} {2020})}\BibitemShut {NoStop}%
\bibitem [{\citenamefont {Indriolo}\ \emph {et~al.}(2015)\citenamefont
  {Indriolo}, \citenamefont {Neufeld}, \citenamefont {Gerin}, \citenamefont
  {Schilke}, \citenamefont {Benz}, \citenamefont {Winkel}, \citenamefont
  {Menten}, \citenamefont {Chambers}, \citenamefont {Black}, \citenamefont
  {Bruderer}, \citenamefont {Falgarone}, \citenamefont {Godard}, \citenamefont
  {Goicoechea}, \citenamefont {Gupta}, \citenamefont {Lis}, \citenamefont
  {Ossenkopf}, \citenamefont {Persson}, \citenamefont {Sonnentrucker},
  \citenamefont {van~der Tak}, \citenamefont {van Dishoeck}, \citenamefont
  {Wolfire},\ and\ \citenamefont {Wyrowski}}]{Indriolo_2015}%
  \BibitemOpen
  \bibfield  {author} {\bibinfo {author} {\bibfnamefont {N.}~\bibnamefont
  {Indriolo}}, \bibinfo {author} {\bibfnamefont {D.~A.}\ \bibnamefont
  {Neufeld}}, \bibinfo {author} {\bibfnamefont {M.}~\bibnamefont {Gerin}},
  \bibinfo {author} {\bibfnamefont {P.}~\bibnamefont {Schilke}}, \bibinfo
  {author} {\bibfnamefont {A.~O.}\ \bibnamefont {Benz}}, \bibinfo {author}
  {\bibfnamefont {B.}~\bibnamefont {Winkel}}, \bibinfo {author} {\bibfnamefont
  {K.~M.}\ \bibnamefont {Menten}}, \bibinfo {author} {\bibfnamefont {E.~T.}\
  \bibnamefont {Chambers}}, \bibinfo {author} {\bibfnamefont {J.~H.}\
  \bibnamefont {Black}}, \bibinfo {author} {\bibfnamefont {S.}~\bibnamefont
  {Bruderer}}, \bibinfo {author} {\bibfnamefont {E.}~\bibnamefont {Falgarone}},
  \bibinfo {author} {\bibfnamefont {B.}~\bibnamefont {Godard}}, \bibinfo
  {author} {\bibfnamefont {J.~R.}\ \bibnamefont {Goicoechea}}, \bibinfo
  {author} {\bibfnamefont {H.}~\bibnamefont {Gupta}}, \bibinfo {author}
  {\bibfnamefont {D.~C.}\ \bibnamefont {Lis}}, \bibinfo {author} {\bibfnamefont
  {V.}~\bibnamefont {Ossenkopf}}, \bibinfo {author} {\bibfnamefont {C.~M.}\
  \bibnamefont {Persson}}, \bibinfo {author} {\bibfnamefont {P.}~\bibnamefont
  {Sonnentrucker}}, \bibinfo {author} {\bibfnamefont {F.~F.~S.}\ \bibnamefont
  {van~der Tak}}, \bibinfo {author} {\bibfnamefont {E.~F.}\ \bibnamefont {van
  Dishoeck}}, \bibinfo {author} {\bibfnamefont {M.~G.}\ \bibnamefont
  {Wolfire}}, \ and\ \bibinfo {author} {\bibfnamefont {F.}~\bibnamefont
  {Wyrowski}},\ }\href {\doibase 10.1088/0004-637X/800/1/40} {\bibfield
  {journal} {\bibinfo  {journal} {The Astrophysical Journal}\ }\textbf
  {\bibinfo {volume} {800}},\ \bibinfo {pages} {40} (\bibinfo {year}
  {2015})}\BibitemShut {NoStop}%
\bibitem [{\citenamefont {{Rivilla}}\ \emph {et~al.}(2022)\citenamefont
  {{Rivilla}}, \citenamefont {{Garc{\'\i}a De La Concepci{\'o}n}},
  \citenamefont {{Jim{\'e}nez-Serra}}, \citenamefont {{Mart{\'\i}n-Pintado}},
  \citenamefont {{Colzi}}, \citenamefont {{Tercero}}, \citenamefont
  {{Meg{\'\i}as}}, \citenamefont {{L{\'o}pez-Gallifa}}, \citenamefont
  {{Mart{\'\i}nez-Henares}}, \citenamefont {{Massalkhi}}, \citenamefont
  {{Mart{\'\i}n}}, \citenamefont {{Zeng}}, \citenamefont {{De Vicente}},
  \citenamefont {{Rico-Villas}}, \citenamefont {{Requena-Torres}},\ and\
  \citenamefont {{Cosentino}}}]{2022FrASS...9.9288R}%
  \BibitemOpen
  \bibfield  {author} {\bibinfo {author} {\bibfnamefont {V.~M.}\ \bibnamefont
  {{Rivilla}}}, \bibinfo {author} {\bibfnamefont {J.}~\bibnamefont
  {{Garc{\'\i}a De La Concepci{\'o}n}}}, \bibinfo {author} {\bibfnamefont
  {I.}~\bibnamefont {{Jim{\'e}nez-Serra}}}, \bibinfo {author} {\bibfnamefont
  {J.}~\bibnamefont {{Mart{\'\i}n-Pintado}}}, \bibinfo {author} {\bibfnamefont
  {L.}~\bibnamefont {{Colzi}}}, \bibinfo {author} {\bibfnamefont
  {B.}~\bibnamefont {{Tercero}}}, \bibinfo {author} {\bibfnamefont
  {A.}~\bibnamefont {{Meg{\'\i}as}}}, \bibinfo {author} {\bibfnamefont
  {{\'A}.}~\bibnamefont {{L{\'o}pez-Gallifa}}}, \bibinfo {author}
  {\bibfnamefont {A.}~\bibnamefont {{Mart{\'\i}nez-Henares}}}, \bibinfo
  {author} {\bibfnamefont {S.}~\bibnamefont {{Massalkhi}}}, \bibinfo {author}
  {\bibfnamefont {S.}~\bibnamefont {{Mart{\'\i}n}}}, \bibinfo {author}
  {\bibfnamefont {S.}~\bibnamefont {{Zeng}}}, \bibinfo {author} {\bibfnamefont
  {P.}~\bibnamefont {{De Vicente}}}, \bibinfo {author} {\bibfnamefont
  {F.}~\bibnamefont {{Rico-Villas}}}, \bibinfo {author} {\bibfnamefont {M.~A.}\
  \bibnamefont {{Requena-Torres}}}, \ and\ \bibinfo {author} {\bibfnamefont
  {G.}~\bibnamefont {{Cosentino}}},\ }\href {\doibase
  10.3389/fspas.2022.829288} {\bibfield  {journal} {\bibinfo  {journal}
  {Frontiers in Astronomy and Space Sciences}\ }\textbf {\bibinfo {volume}
  {9}},\ \bibinfo {eid} {829288} (\bibinfo {year} {2022})},\ \Eprint
  {http://arxiv.org/abs/2202.13928} {arXiv:2202.13928 [astro-ph.GA]}
  \BibitemShut {NoStop}%
\bibitem [{\citenamefont {Yusef-Zadeh}\ \emph {et~al.}(2012)\citenamefont
  {Yusef-Zadeh}, \citenamefont {Hewitt}, \citenamefont {Wardle}, \citenamefont
  {Tatischeff}, \citenamefont {Roberts}, \citenamefont {Cotton}, \citenamefont
  {Uchiyama}, \citenamefont {Nobukawa}, \citenamefont {Tsuru}, \citenamefont
  {Heinke},\ and\ \citenamefont {Royster}}]{Yusef_Zadeh_2012}%
  \BibitemOpen
  \bibfield  {author} {\bibinfo {author} {\bibfnamefont {F.}~\bibnamefont
  {Yusef-Zadeh}}, \bibinfo {author} {\bibfnamefont {J.~W.}\ \bibnamefont
  {Hewitt}}, \bibinfo {author} {\bibfnamefont {M.}~\bibnamefont {Wardle}},
  \bibinfo {author} {\bibfnamefont {V.}~\bibnamefont {Tatischeff}}, \bibinfo
  {author} {\bibfnamefont {D.~A.}\ \bibnamefont {Roberts}}, \bibinfo {author}
  {\bibfnamefont {W.}~\bibnamefont {Cotton}}, \bibinfo {author} {\bibfnamefont
  {H.}~\bibnamefont {Uchiyama}}, \bibinfo {author} {\bibfnamefont
  {M.}~\bibnamefont {Nobukawa}}, \bibinfo {author} {\bibfnamefont {T.~G.}\
  \bibnamefont {Tsuru}}, \bibinfo {author} {\bibfnamefont {C.}~\bibnamefont
  {Heinke}}, \ and\ \bibinfo {author} {\bibfnamefont {M.}~\bibnamefont
  {Royster}},\ }\href {\doibase 10.1088/0004-637x/762/1/33} {\bibfield
  {journal} {\bibinfo  {journal} {The Astrophysical Journal}\ }\textbf
  {\bibinfo {volume} {762}},\ \bibinfo {pages} {33} (\bibinfo {year}
  {2012})}\BibitemShut {NoStop}%
\bibitem [{\citenamefont {{Yusef-Zadeh}}\ \emph {et~al.}(2007)\citenamefont
  {{Yusef-Zadeh}}, \citenamefont {{Muno}}, \citenamefont {{Wardle}},\ and\
  \citenamefont {{Lis}}}]{2007ApJ...656..847Y}%
  \BibitemOpen
  \bibfield  {author} {\bibinfo {author} {\bibfnamefont {F.}~\bibnamefont
  {{Yusef-Zadeh}}}, \bibinfo {author} {\bibfnamefont {M.}~\bibnamefont
  {{Muno}}}, \bibinfo {author} {\bibfnamefont {M.}~\bibnamefont {{Wardle}}}, \
  and\ \bibinfo {author} {\bibfnamefont {D.~C.}\ \bibnamefont {{Lis}}},\ }\href
  {\doibase 10.1086/510663} {\bibfield  {journal} {\bibinfo  {journal} {\apj}\
  }\textbf {\bibinfo {volume} {656}},\ \bibinfo {pages} {847} (\bibinfo {year}
  {2007})},\ \Eprint {http://arxiv.org/abs/astro-ph/0608710}
  {arXiv:astro-ph/0608710 [astro-ph]} \BibitemShut {NoStop}%
\bibitem [{\citenamefont {Oka}\ \emph {et~al.}(2005)\citenamefont {Oka},
  \citenamefont {Geballe}, \citenamefont {Goto}, \citenamefont {Usuda},\ and\
  \citenamefont {McCall}}]{Oka_2005}%
  \BibitemOpen
  \bibfield  {author} {\bibinfo {author} {\bibfnamefont {T.}~\bibnamefont
  {Oka}}, \bibinfo {author} {\bibfnamefont {T.~R.}\ \bibnamefont {Geballe}},
  \bibinfo {author} {\bibfnamefont {M.}~\bibnamefont {Goto}}, \bibinfo {author}
  {\bibfnamefont {T.}~\bibnamefont {Usuda}}, \ and\ \bibinfo {author}
  {\bibfnamefont {B.~J.}\ \bibnamefont {McCall}},\ }\href {\doibase
  10.1086/432679} {\bibfield  {journal} {\bibinfo  {journal} {The Astrophysical
  Journal}\ }\textbf {\bibinfo {volume} {632}},\ \bibinfo {pages} {882–893}
  (\bibinfo {year} {2005})}\BibitemShut {NoStop}%
\bibitem [{\citenamefont {Goto}\ \emph {et~al.}(2011)\citenamefont {Goto},
  \citenamefont {Usuda}, \citenamefont {Geballe}, \citenamefont {Indriolo},
  \citenamefont {Mccall}, \citenamefont {Henning},\ and\ \citenamefont
  {Oka}}]{Goto_2011}%
  \BibitemOpen
  \bibfield  {author} {\bibinfo {author} {\bibfnamefont {M.}~\bibnamefont
  {Goto}}, \bibinfo {author} {\bibfnamefont {T.}~\bibnamefont {Usuda}},
  \bibinfo {author} {\bibfnamefont {T.~R.}\ \bibnamefont {Geballe}}, \bibinfo
  {author} {\bibfnamefont {N.}~\bibnamefont {Indriolo}}, \bibinfo {author}
  {\bibfnamefont {B.~J.}\ \bibnamefont {Mccall}}, \bibinfo {author}
  {\bibfnamefont {T.}~\bibnamefont {Henning}}, \ and\ \bibinfo {author}
  {\bibfnamefont {T.}~\bibnamefont {Oka}},\ }\href {\doibase
  10.1093/pasj/63.2.l13} {\bibfield  {journal} {\bibinfo  {journal}
  {Publications of the Astronomical Society of Japan}\ }\textbf {\bibinfo
  {volume} {63}},\ \bibinfo {pages} {L13–L17} (\bibinfo {year}
  {2011})}\BibitemShut {NoStop}%
\bibitem [{\citenamefont {Geballe}\ and\ \citenamefont
  {Oka}(2010)}]{Geballe_2010}%
  \BibitemOpen
  \bibfield  {author} {\bibinfo {author} {\bibfnamefont {T.~R.}\ \bibnamefont
  {Geballe}}\ and\ \bibinfo {author} {\bibfnamefont {T.}~\bibnamefont {Oka}},\
  }\href {\doibase 10.1088/2041-8205/709/1/L70} {\bibfield  {journal} {\bibinfo
   {journal} {The Astrophysical Journal Letters}\ }\textbf {\bibinfo {volume}
  {709}},\ \bibinfo {pages} {L70} (\bibinfo {year} {2010})}\BibitemShut
  {NoStop}%
\bibitem [{\citenamefont {{Le Petit}}\ \emph {et~al.}(2016)\citenamefont {{Le
  Petit}}, \citenamefont {{Ruaud}}, \citenamefont {{Bron}}, \citenamefont
  {{Godard}}, \citenamefont {{Roueff}}, \citenamefont {{Languignon}},\ and\
  \citenamefont {{Le Bourlot}}}]{IoRPetit}%
  \BibitemOpen
  \bibfield  {author} {\bibinfo {author} {\bibfnamefont {F.}~\bibnamefont {{Le
  Petit}}}, \bibinfo {author} {\bibfnamefont {M.}~\bibnamefont {{Ruaud}}},
  \bibinfo {author} {\bibfnamefont {E.}~\bibnamefont {{Bron}}}, \bibinfo
  {author} {\bibfnamefont {B.}~\bibnamefont {{Godard}}}, \bibinfo {author}
  {\bibfnamefont {E.}~\bibnamefont {{Roueff}}}, \bibinfo {author}
  {\bibfnamefont {D.}~\bibnamefont {{Languignon}}}, \ and\ \bibinfo {author}
  {\bibfnamefont {J.}~\bibnamefont {{Le Bourlot}}},\ }\href {\doibase
  10.1051/0004-6361/201526658} {\bibfield  {journal} {\bibinfo  {journal}
  {Astronomy \& Astrophysics}\ }\textbf {\bibinfo {volume} {585}},\ \bibinfo
  {eid} {A105} (\bibinfo {year} {2016})},\ \Eprint
  {http://arxiv.org/abs/1510.02221} {arXiv:1510.02221 [astro-ph.GA]}
  \BibitemShut {NoStop}%
\bibitem [{\citenamefont {Sanz-Novo}\ \emph {et~al.}(2024)\citenamefont
  {Sanz-Novo}, \citenamefont {Rivilla}, \citenamefont {Jiménez-Serra},
  \citenamefont {Martín-Pintado}, \citenamefont {Colzi}, \citenamefont {Zeng},
  \citenamefont {Megías}, \citenamefont {Álvaro López-Gallifa},
  \citenamefont {Martínez-Henares}, \citenamefont {Massalkhi}, \citenamefont
  {Tercero}, \citenamefont {de~Vicente}, \citenamefont {Andrés}, \citenamefont
  {Martín},\ and\ \citenamefont {Requena-Torres}}]{Sanz-Novo_2024}%
  \BibitemOpen
  \bibfield  {author} {\bibinfo {author} {\bibfnamefont {M.}~\bibnamefont
  {Sanz-Novo}}, \bibinfo {author} {\bibfnamefont {V.~M.}\ \bibnamefont
  {Rivilla}}, \bibinfo {author} {\bibfnamefont {I.}~\bibnamefont
  {Jiménez-Serra}}, \bibinfo {author} {\bibfnamefont {J.}~\bibnamefont
  {Martín-Pintado}}, \bibinfo {author} {\bibfnamefont {L.}~\bibnamefont
  {Colzi}}, \bibinfo {author} {\bibfnamefont {S.}~\bibnamefont {Zeng}},
  \bibinfo {author} {\bibfnamefont {A.}~\bibnamefont {Megías}}, \bibinfo
  {author} {\bibnamefont {Álvaro López-Gallifa}}, \bibinfo {author}
  {\bibfnamefont {A.}~\bibnamefont {Martínez-Henares}}, \bibinfo {author}
  {\bibfnamefont {S.}~\bibnamefont {Massalkhi}}, \bibinfo {author}
  {\bibfnamefont {B.}~\bibnamefont {Tercero}}, \bibinfo {author} {\bibfnamefont
  {P.}~\bibnamefont {de~Vicente}}, \bibinfo {author} {\bibfnamefont {D.~S.}\
  \bibnamefont {Andrés}}, \bibinfo {author} {\bibfnamefont {S.}~\bibnamefont
  {Martín}}, \ and\ \bibinfo {author} {\bibfnamefont {M.~A.}\ \bibnamefont
  {Requena-Torres}},\ }\href {\doibase 10.3847/1538-4357/ad2c01} {\bibfield
  {journal} {\bibinfo  {journal} {The Astrophysical Journal}\ }\textbf
  {\bibinfo {volume} {965}},\ \bibinfo {pages} {149} (\bibinfo {year}
  {2024})}\BibitemShut {NoStop}%
\bibitem [{\citenamefont {Coutens}\ \emph {et~al.}(2017)\citenamefont
  {Coutens}, \citenamefont {Rawlings}, \citenamefont {Viti},\ and\
  \citenamefont {Williams}}]{IoRMethanol}%
  \BibitemOpen
  \bibfield  {author} {\bibinfo {author} {\bibfnamefont {A.}~\bibnamefont
  {Coutens}}, \bibinfo {author} {\bibfnamefont {J.~M.~C.}\ \bibnamefont
  {Rawlings}}, \bibinfo {author} {\bibfnamefont {S.}~\bibnamefont {Viti}}, \
  and\ \bibinfo {author} {\bibfnamefont {D.~A.}\ \bibnamefont {Williams}},\
  }\href {\doibase 10.1093/mnras/stx119} {\bibfield  {journal} {\bibinfo
  {journal} {Monthly Notices of the Royal Astronomical Society}\ }\textbf
  {\bibinfo {volume} {467}},\ \bibinfo {pages} {737} (\bibinfo {year}
  {2017})},\ \Eprint
  {http://arxiv.org/abs/https://academic.oup.com/mnras/article-pdf/467/1/737/10493726/stx119.pdf}
  {https://academic.oup.com/mnras/article-pdf/467/1/737/10493726/stx119.pdf}
  \BibitemShut {NoStop}%
\bibitem [{\citenamefont {Oka}\ and\ \citenamefont {Geballe}(2020)}]{Oka_2020}%
  \BibitemOpen
  \bibfield  {author} {\bibinfo {author} {\bibfnamefont {T.}~\bibnamefont
  {Oka}}\ and\ \bibinfo {author} {\bibfnamefont {T.~R.}\ \bibnamefont
  {Geballe}},\ }\href {\doibase 10.3847/1538-4357/abb1b5} {\bibfield  {journal}
  {\bibinfo  {journal} {The Astrophysical Journal}\ }\textbf {\bibinfo {volume}
  {902}},\ \bibinfo {pages} {9} (\bibinfo {year} {2020})}\BibitemShut {NoStop}%
\bibitem [{\citenamefont {Indriolo}\ \emph {et~al.}(2009)\citenamefont
  {Indriolo}, \citenamefont {Fields},\ and\ \citenamefont
  {McCall}}]{Indriolo_2009}%
  \BibitemOpen
  \bibfield  {author} {\bibinfo {author} {\bibfnamefont {N.}~\bibnamefont
  {Indriolo}}, \bibinfo {author} {\bibfnamefont {B.~D.}\ \bibnamefont
  {Fields}}, \ and\ \bibinfo {author} {\bibfnamefont {B.~J.}\ \bibnamefont
  {McCall}},\ }\href {\doibase 10.1088/0004-637X/694/1/257} {\bibfield
  {journal} {\bibinfo  {journal} {The Astrophysical Journal}\ }\textbf
  {\bibinfo {volume} {694}},\ \bibinfo {pages} {257} (\bibinfo {year}
  {2009})}\BibitemShut {NoStop}%
\bibitem [{\citenamefont {Goto}\ \emph {et~al.}(2014)\citenamefont {Goto},
  \citenamefont {Geballe}, \citenamefont {Indriolo}, \citenamefont
  {Yusef-Zadeh}, \citenamefont {Usuda}, \citenamefont {Henning},\ and\
  \citenamefont {Oka}}]{Goto_2014}%
  \BibitemOpen
  \bibfield  {author} {\bibinfo {author} {\bibfnamefont {M.}~\bibnamefont
  {Goto}}, \bibinfo {author} {\bibfnamefont {T.~R.}\ \bibnamefont {Geballe}},
  \bibinfo {author} {\bibfnamefont {N.}~\bibnamefont {Indriolo}}, \bibinfo
  {author} {\bibfnamefont {F.}~\bibnamefont {Yusef-Zadeh}}, \bibinfo {author}
  {\bibfnamefont {T.}~\bibnamefont {Usuda}}, \bibinfo {author} {\bibfnamefont
  {T.}~\bibnamefont {Henning}}, \ and\ \bibinfo {author} {\bibfnamefont
  {T.}~\bibnamefont {Oka}},\ }\href {\doibase 10.1088/0004-637X/786/2/96}
  {\bibfield  {journal} {\bibinfo  {journal} {The Astrophysical Journal}\
  }\textbf {\bibinfo {volume} {786}},\ \bibinfo {pages} {96} (\bibinfo {year}
  {2014})}\BibitemShut {NoStop}%
\bibitem [{\citenamefont {van~der Tak}\ \emph {et~al.}(2006)\citenamefont
  {van~der Tak}, \citenamefont {Belloche}, \citenamefont {Schilke},
  \citenamefont {Güsten}, \citenamefont {Philipp}, \citenamefont {Comito},
  \citenamefont {Bergman},\ and\ \citenamefont {Nyman}}]{van_der_Tak_2006}%
  \BibitemOpen
  \bibfield  {author} {\bibinfo {author} {\bibfnamefont {F.~F.~S.}\
  \bibnamefont {van~der Tak}}, \bibinfo {author} {\bibfnamefont
  {A.}~\bibnamefont {Belloche}}, \bibinfo {author} {\bibfnamefont
  {P.}~\bibnamefont {Schilke}}, \bibinfo {author} {\bibfnamefont
  {R.}~\bibnamefont {Güsten}}, \bibinfo {author} {\bibfnamefont
  {S.}~\bibnamefont {Philipp}}, \bibinfo {author} {\bibfnamefont
  {C.}~\bibnamefont {Comito}}, \bibinfo {author} {\bibfnamefont
  {P.}~\bibnamefont {Bergman}}, \ and\ \bibinfo {author} {\bibfnamefont
  {L.-A.}\ \bibnamefont {Nyman}},\ }\href {\doibase 10.1051/0004-6361:20065289}
  {\bibfield  {journal} {\bibinfo  {journal} {Astronomy \& Astrophysics}\
  }\textbf {\bibinfo {volume} {454}},\ \bibinfo {pages} {L99–L102} (\bibinfo
  {year} {2006})}\BibitemShut {NoStop}%
\bibitem [{\citenamefont {Phan}\ \emph {et~al.}(2018)\citenamefont {Phan},
  \citenamefont {Morlino},\ and\ \citenamefont {Gabici}}]{Phan_2018}%
  \BibitemOpen
  \bibfield  {author} {\bibinfo {author} {\bibfnamefont {V.~H.~M.}\
  \bibnamefont {Phan}}, \bibinfo {author} {\bibfnamefont {G.}~\bibnamefont
  {Morlino}}, \ and\ \bibinfo {author} {\bibfnamefont {S.}~\bibnamefont
  {Gabici}},\ }\href {\doibase 10.1093/mnras/sty2235} {\bibfield  {journal}
  {\bibinfo  {journal} {Monthly Notices of the Royal Astronomical Society}\ }
  (\bibinfo {year} {2018}),\ 10.1093/mnras/sty2235}\BibitemShut {NoStop}%
\bibitem [{\citenamefont {Boudaud}\ \emph {et~al.}(2017)\citenamefont
  {Boudaud}, \citenamefont {Lavalle},\ and\ \citenamefont
  {Salati}}]{Boudaud:2016mos}%
  \BibitemOpen
  \bibfield  {author} {\bibinfo {author} {\bibfnamefont {M.}~\bibnamefont
  {Boudaud}}, \bibinfo {author} {\bibfnamefont {J.}~\bibnamefont {Lavalle}}, \
  and\ \bibinfo {author} {\bibfnamefont {P.}~\bibnamefont {Salati}},\ }\href
  {\doibase 10.1103/PhysRevLett.119.021103} {\bibfield  {journal} {\bibinfo
  {journal} {Phys. Rev. Lett.}\ }\textbf {\bibinfo {volume} {119}},\ \bibinfo
  {pages} {021103} (\bibinfo {year} {2017})},\ \Eprint
  {http://arxiv.org/abs/1612.07698} {arXiv:1612.07698 [astro-ph.HE]}
  \BibitemShut {NoStop}%
\bibitem [{\citenamefont {Cirelli}\ \emph
  {et~al.}(2023{\natexlab{a}})\citenamefont {Cirelli}, \citenamefont
  {Fornengo}, \citenamefont {Koechler}, \citenamefont {Pinetti},\ and\
  \citenamefont {Roach}}]{Cirelli:2023tnx}%
  \BibitemOpen
  \bibfield  {author} {\bibinfo {author} {\bibfnamefont {M.}~\bibnamefont
  {Cirelli}}, \bibinfo {author} {\bibfnamefont {N.}~\bibnamefont {Fornengo}},
  \bibinfo {author} {\bibfnamefont {J.}~\bibnamefont {Koechler}}, \bibinfo
  {author} {\bibfnamefont {E.}~\bibnamefont {Pinetti}}, \ and\ \bibinfo
  {author} {\bibfnamefont {B.~M.}\ \bibnamefont {Roach}},\ }\href {\doibase
  10.1088/1475-7516/2023/07/026} {\bibfield  {journal} {\bibinfo  {journal}
  {JCAP}\ }\textbf {\bibinfo {volume} {07}},\ \bibinfo {pages} {026} (\bibinfo
  {year} {2023}{\natexlab{a}})},\ \Eprint {http://arxiv.org/abs/2303.08854}
  {arXiv:2303.08854 [hep-ph]} \BibitemShut {NoStop}%
\bibitem [{\citenamefont {Cirelli}\ \emph {et~al.}(2024)\citenamefont
  {Cirelli}, \citenamefont {Strumia},\ and\ \citenamefont
  {Zupan}}]{Cirelli:2024ssz}%
  \BibitemOpen
  \bibfield  {author} {\bibinfo {author} {\bibfnamefont {M.}~\bibnamefont
  {Cirelli}}, \bibinfo {author} {\bibfnamefont {A.}~\bibnamefont {Strumia}}, \
  and\ \bibinfo {author} {\bibfnamefont {J.}~\bibnamefont {Zupan}},\
  }\href@noop {} {\  (\bibinfo {year} {2024})},\ \Eprint
  {http://arxiv.org/abs/2406.01705} {arXiv:2406.01705 [hep-ph]} \BibitemShut
  {NoStop}%
\bibitem [{\citenamefont {Battaglieri}\ \emph {et~al.}(2017)\citenamefont
  {Battaglieri} \emph {et~al.}}]{Battaglieri:2017aum}%
  \BibitemOpen
  \bibfield  {author} {\bibinfo {author} {\bibfnamefont {M.}~\bibnamefont
  {Battaglieri}} \emph {et~al.},\ }in\ \href@noop {} {\emph {\bibinfo
  {booktitle} {{U.S. Cosmic Visions: New Ideas in Dark Matter}}}}\ (\bibinfo
  {year} {2017})\ \Eprint {http://arxiv.org/abs/1707.04591} {arXiv:1707.04591
  [hep-ph]} \BibitemShut {NoStop}%
\bibitem [{\citenamefont {Cirelli}\ \emph {et~al.}(2021)\citenamefont
  {Cirelli}, \citenamefont {Fornengo}, \citenamefont {Kavanagh},\ and\
  \citenamefont {Pinetti}}]{Cirelli:2020bpc}%
  \BibitemOpen
  \bibfield  {author} {\bibinfo {author} {\bibfnamefont {M.}~\bibnamefont
  {Cirelli}}, \bibinfo {author} {\bibfnamefont {N.}~\bibnamefont {Fornengo}},
  \bibinfo {author} {\bibfnamefont {B.~J.}\ \bibnamefont {Kavanagh}}, \ and\
  \bibinfo {author} {\bibfnamefont {E.}~\bibnamefont {Pinetti}},\ }\href
  {\doibase 10.1103/PhysRevD.103.063022} {\bibfield  {journal} {\bibinfo
  {journal} {Phys. Rev. D}\ }\textbf {\bibinfo {volume} {103}},\ \bibinfo
  {pages} {063022} (\bibinfo {year} {2021})},\ \Eprint
  {http://arxiv.org/abs/2007.11493} {arXiv:2007.11493 [hep-ph]} \BibitemShut
  {NoStop}%
\bibitem [{\citenamefont {Balan}\ \emph
  {et~al.}(2024{\natexlab{a}})\citenamefont {Balan} \emph
  {et~al.}}]{Balan:2024cmq}%
  \BibitemOpen
  \bibfield  {author} {\bibinfo {author} {\bibfnamefont {S.}~\bibnamefont
  {Balan}} \emph {et~al.},\ }\href@noop {} {\  (\bibinfo {year}
  {2024}{\natexlab{a}})},\ \Eprint {http://arxiv.org/abs/2405.17548}
  {arXiv:2405.17548 [hep-ph]} \BibitemShut {NoStop}%
\bibitem [{\citenamefont {Navarro}\ \emph {et~al.}(1996)\citenamefont
  {Navarro}, \citenamefont {Frenk},\ and\ \citenamefont
  {White}}]{Navarro:1995iw}%
  \BibitemOpen
  \bibfield  {author} {\bibinfo {author} {\bibfnamefont {J.~F.}\ \bibnamefont
  {Navarro}}, \bibinfo {author} {\bibfnamefont {C.~S.}\ \bibnamefont {Frenk}},
  \ and\ \bibinfo {author} {\bibfnamefont {S.~D.~M.}\ \bibnamefont {White}},\
  }\href {\doibase 10.1086/177173} {\bibfield  {journal} {\bibinfo  {journal}
  {Astrophys. J.}\ }\textbf {\bibinfo {volume} {462}},\ \bibinfo {pages} {563}
  (\bibinfo {year} {1996})},\ \Eprint {http://arxiv.org/abs/astro-ph/9508025}
  {arXiv:astro-ph/9508025} \BibitemShut {NoStop}%
\bibitem [{\citenamefont {Prantzos}\ \emph {et~al.}(2011)\citenamefont
  {Prantzos} \emph {et~al.}}]{Prantzos:2010wi}%
  \BibitemOpen
  \bibfield  {author} {\bibinfo {author} {\bibfnamefont {N.}~\bibnamefont
  {Prantzos}} \emph {et~al.},\ }\href {\doibase 10.1103/RevModPhys.83.1001}
  {\bibfield  {journal} {\bibinfo  {journal} {Rev. Mod. Phys.}\ }\textbf
  {\bibinfo {volume} {83}},\ \bibinfo {pages} {1001} (\bibinfo {year}
  {2011})},\ \Eprint {http://arxiv.org/abs/1009.4620} {arXiv:1009.4620
  [astro-ph.HE]} \BibitemShut {NoStop}%
\bibitem [{\citenamefont {Kierans}\ \emph {et~al.}(2019)\citenamefont {Kierans}
  \emph {et~al.}}]{kierans2019positron}%
  \BibitemOpen
  \bibfield  {author} {\bibinfo {author} {\bibfnamefont {C.~A.}\ \bibnamefont
  {Kierans}} \emph {et~al.},\ }\href@noop {} {\enquote {\bibinfo {title}
  {Positron annihilation in the galaxy},}\ } (\bibinfo {year} {2019}),\ \Eprint
  {http://arxiv.org/abs/1903.05569} {arXiv:1903.05569 [astro-ph.HE]}
  \BibitemShut {NoStop}%
\bibitem [{\citenamefont {Siegert}(2023)}]{Siegert_2023}%
  \BibitemOpen
  \bibfield  {author} {\bibinfo {author} {\bibfnamefont {T.}~\bibnamefont
  {Siegert}},\ }\href {\doibase 10.1007/s10509-023-04184-4} {\bibfield
  {journal} {\bibinfo  {journal} {Astrophysics and Space Science}\ }\textbf
  {\bibinfo {volume} {368}} (\bibinfo {year} {2023}),\
  10.1007/s10509-023-04184-4}\BibitemShut {NoStop}%
\bibitem [{\citenamefont {Boehm}\ \emph {et~al.}(2004)\citenamefont {Boehm},
  \citenamefont {Hooper}, \citenamefont {Silk}, \citenamefont {Casse},\ and\
  \citenamefont {Paul}}]{Boehm}%
  \BibitemOpen
  \bibfield  {author} {\bibinfo {author} {\bibfnamefont {C.}~\bibnamefont
  {Boehm}}, \bibinfo {author} {\bibfnamefont {D.}~\bibnamefont {Hooper}},
  \bibinfo {author} {\bibfnamefont {J.}~\bibnamefont {Silk}}, \bibinfo {author}
  {\bibfnamefont {M.}~\bibnamefont {Casse}}, \ and\ \bibinfo {author}
  {\bibfnamefont {J.}~\bibnamefont {Paul}},\ }\href {\doibase
  10.1103/PhysRevLett.92.101301} {\bibfield  {journal} {\bibinfo  {journal}
  {Phys. Rev. Lett.}\ }\textbf {\bibinfo {volume} {92}},\ \bibinfo {pages}
  {101301} (\bibinfo {year} {2004})}\BibitemShut {NoStop}%
\bibitem [{\citenamefont {De~la Torre~Luque}\ \emph
  {et~al.}(2023{\natexlab{a}})\citenamefont {De~la Torre~Luque}, \citenamefont
  {Balaji},\ and\ \citenamefont {Silk}}]{DelaTorreLuque:2023cef}%
  \BibitemOpen
  \bibfield  {author} {\bibinfo {author} {\bibfnamefont {P.}~\bibnamefont
  {De~la Torre~Luque}}, \bibinfo {author} {\bibfnamefont {S.}~\bibnamefont
  {Balaji}}, \ and\ \bibinfo {author} {\bibfnamefont {J.}~\bibnamefont
  {Silk}},\ }\href@noop {} {\bibfield  {journal} {\bibinfo  {journal} {{}}\ }
  (\bibinfo {year} {2023}{\natexlab{a}})},\ \Eprint
  {http://arxiv.org/abs/2312.04907} {arXiv:2312.04907 [hep-ph]} \BibitemShut
  {NoStop}%
\bibitem [{\citenamefont {Vincent}\ \emph {et~al.}(2012)\citenamefont
  {Vincent}, \citenamefont {Martin},\ and\ \citenamefont
  {Cline}}]{Vincent_2012}%
  \BibitemOpen
  \bibfield  {author} {\bibinfo {author} {\bibfnamefont {A.~C.}\ \bibnamefont
  {Vincent}}, \bibinfo {author} {\bibfnamefont {P.}~\bibnamefont {Martin}}, \
  and\ \bibinfo {author} {\bibfnamefont {J.~M.}\ \bibnamefont {Cline}},\ }\href
  {\doibase 10.1088/1475-7516/2012/04/022} {\bibfield  {journal} {\bibinfo
  {journal} {Journal of Cosmology and Astroparticle Physics}\ }\textbf
  {\bibinfo {volume} {2012}},\ \bibinfo {pages} {022} (\bibinfo {year}
  {2012})}\BibitemShut {NoStop}%
\bibitem [{\citenamefont {Siegert}\ \emph {et~al.}(2016)\citenamefont
  {Siegert}, \citenamefont {Diehl}, \citenamefont {Khachatryan}, \citenamefont
  {Krause}, \citenamefont {Guglielmetti}, \citenamefont {Greiner},
  \citenamefont {Strong},\ and\ \citenamefont {Zhang}}]{Siegert:2015knp}%
  \BibitemOpen
  \bibfield  {author} {\bibinfo {author} {\bibfnamefont {T.}~\bibnamefont
  {Siegert}}, \bibinfo {author} {\bibfnamefont {R.}~\bibnamefont {Diehl}},
  \bibinfo {author} {\bibfnamefont {G.}~\bibnamefont {Khachatryan}}, \bibinfo
  {author} {\bibfnamefont {M.~G.~H.}\ \bibnamefont {Krause}}, \bibinfo {author}
  {\bibfnamefont {F.}~\bibnamefont {Guglielmetti}}, \bibinfo {author}
  {\bibfnamefont {J.}~\bibnamefont {Greiner}}, \bibinfo {author} {\bibfnamefont
  {A.~W.}\ \bibnamefont {Strong}}, \ and\ \bibinfo {author} {\bibfnamefont
  {X.}~\bibnamefont {Zhang}},\ }\href {\doibase 10.1051/0004-6361/201527510}
  {\bibfield  {journal} {\bibinfo  {journal} {Astron. Astrophys.}\ }\textbf
  {\bibinfo {volume} {586}},\ \bibinfo {pages} {A84} (\bibinfo {year}
  {2016})},\ \Eprint {http://arxiv.org/abs/1512.00325} {arXiv:1512.00325
  [astro-ph.HE]} \BibitemShut {NoStop}%
\bibitem [{\citenamefont {Kierans}\ \emph {et~al.}(2020)\citenamefont {Kierans}
  \emph {et~al.}}]{Kierans:2019aqz}%
  \BibitemOpen
  \bibfield  {author} {\bibinfo {author} {\bibfnamefont {C.~A.}\ \bibnamefont
  {Kierans}} \emph {et~al.},\ }\href {\doibase 10.3847/1538-4357/ab89a9}
  {\bibfield  {journal} {\bibinfo  {journal} {Astrophys. J.}\ }\textbf
  {\bibinfo {volume} {895}},\ \bibinfo {pages} {44} (\bibinfo {year} {2020})},\
  \Eprint {http://arxiv.org/abs/1912.00110} {arXiv:1912.00110 [astro-ph.HE]}
  \BibitemShut {NoStop}%
\bibitem [{\citenamefont {Badertscher}\ \emph {et~al.}(2007)\citenamefont
  {Badertscher}, \citenamefont {Crivelli}, \citenamefont {Fetscher},
  \citenamefont {Gendotti}, \citenamefont {Gninenko}, \citenamefont {Postoev},
  \citenamefont {Rubbia}, \citenamefont {Samoylenko},\ and\ \citenamefont
  {Sillou}}]{Badertscher_2007}%
  \BibitemOpen
  \bibfield  {author} {\bibinfo {author} {\bibfnamefont {A.}~\bibnamefont
  {Badertscher}}, \bibinfo {author} {\bibfnamefont {P.}~\bibnamefont
  {Crivelli}}, \bibinfo {author} {\bibfnamefont {W.}~\bibnamefont {Fetscher}},
  \bibinfo {author} {\bibfnamefont {U.}~\bibnamefont {Gendotti}}, \bibinfo
  {author} {\bibfnamefont {S.~N.}\ \bibnamefont {Gninenko}}, \bibinfo {author}
  {\bibfnamefont {V.}~\bibnamefont {Postoev}}, \bibinfo {author} {\bibfnamefont
  {A.}~\bibnamefont {Rubbia}}, \bibinfo {author} {\bibfnamefont
  {V.}~\bibnamefont {Samoylenko}}, \ and\ \bibinfo {author} {\bibfnamefont
  {D.}~\bibnamefont {Sillou}},\ }\href {\doibase 10.1103/physrevd.75.032004}
  {\bibfield  {journal} {\bibinfo  {journal} {Physical Review D}\ }\textbf
  {\bibinfo {volume} {75}} (\bibinfo {year} {2007}),\
  10.1103/physrevd.75.032004}\BibitemShut {NoStop}%
\bibitem [{\citenamefont {Ferriere}\ \emph {et~al.}(2007)\citenamefont
  {Ferriere}, \citenamefont {Gillard},\ and\ \citenamefont
  {Jean}}]{ferriere2007spatial}%
  \BibitemOpen
  \bibfield  {author} {\bibinfo {author} {\bibfnamefont {K.}~\bibnamefont
  {Ferriere}}, \bibinfo {author} {\bibfnamefont {W.}~\bibnamefont {Gillard}}, \
  and\ \bibinfo {author} {\bibfnamefont {P.}~\bibnamefont {Jean}},\ }\href@noop
  {} {\bibfield  {journal} {\bibinfo  {journal} {Astronomy \& Astrophysics}\
  }\textbf {\bibinfo {volume} {467}},\ \bibinfo {pages} {611} (\bibinfo {year}
  {2007})}\BibitemShut {NoStop}%
\bibitem [{\citenamefont {Padovani}\ \emph {et~al.}(2009)\citenamefont
  {Padovani}, \citenamefont {Galli},\ and\ \citenamefont
  {Glassgold}}]{Padovani_2009}%
  \BibitemOpen
  \bibfield  {author} {\bibinfo {author} {\bibfnamefont {M.}~\bibnamefont
  {Padovani}}, \bibinfo {author} {\bibfnamefont {D.}~\bibnamefont {Galli}}, \
  and\ \bibinfo {author} {\bibfnamefont {A.~E.}\ \bibnamefont {Glassgold}},\
  }\href {\doibase 10.1051/0004-6361/200911794} {\bibfield  {journal} {\bibinfo
   {journal} {Astronomy \& Astrophysics}\ }\textbf {\bibinfo {volume} {501}},\
  \bibinfo {pages} {619–631} (\bibinfo {year} {2009})}\BibitemShut {NoStop}%
\bibitem [{\citenamefont {Phan}\ \emph {et~al.}(2023)\citenamefont {Phan},
  \citenamefont {Recchia}, \citenamefont {Mertsch},\ and\ \citenamefont
  {Gabici}}]{Phan_2023}%
  \BibitemOpen
  \bibfield  {author} {\bibinfo {author} {\bibfnamefont {V.~H.~M.}\
  \bibnamefont {Phan}}, \bibinfo {author} {\bibfnamefont {S.}~\bibnamefont
  {Recchia}}, \bibinfo {author} {\bibfnamefont {P.}~\bibnamefont {Mertsch}}, \
  and\ \bibinfo {author} {\bibfnamefont {S.}~\bibnamefont {Gabici}},\ }\href
  {\doibase 10.1103/physrevd.107.123006} {\bibfield  {journal} {\bibinfo
  {journal} {Physical Review D}\ }\textbf {\bibinfo {volume} {107}} (\bibinfo
  {year} {2023}),\ 10.1103/physrevd.107.123006}\BibitemShut {NoStop}%
\bibitem [{\citenamefont {Krause}\ \emph {et~al.}(2015)\citenamefont {Krause},
  \citenamefont {Morlino},\ and\ \citenamefont
  {Gabici}}]{krause2015crimecosmicray}%
  \BibitemOpen
  \bibfield  {author} {\bibinfo {author} {\bibfnamefont {J.}~\bibnamefont
  {Krause}}, \bibinfo {author} {\bibfnamefont {G.}~\bibnamefont {Morlino}}, \
  and\ \bibinfo {author} {\bibfnamefont {S.}~\bibnamefont {Gabici}},\ }\href
  {https://arxiv.org/abs/1507.05127} {\enquote {\bibinfo {title} {Crime -
  cosmic ray interactions in molecular environments},}\ } (\bibinfo {year}
  {2015}),\ \Eprint {http://arxiv.org/abs/1507.05127} {arXiv:1507.05127
  [astro-ph.HE]} \BibitemShut {NoStop}%
\bibitem [{\citenamefont {De~la Torre~Luque}\ \emph
  {et~al.}(2023{\natexlab{b}})\citenamefont {De~la Torre~Luque}, \citenamefont
  {Balaji},\ and\ \citenamefont {Carenza}}]{DelaTorreLuque:2023huu}%
  \BibitemOpen
  \bibfield  {author} {\bibinfo {author} {\bibfnamefont {P.}~\bibnamefont
  {De~la Torre~Luque}}, \bibinfo {author} {\bibfnamefont {S.}~\bibnamefont
  {Balaji}}, \ and\ \bibinfo {author} {\bibfnamefont {P.}~\bibnamefont
  {Carenza}},\ }\href@noop {} {\bibfield  {journal} {\bibinfo  {journal}
  {{Accepted in PRD}}\ } (\bibinfo {year} {2023}{\natexlab{b}})},\ \Eprint
  {http://arxiv.org/abs/2307.13731} {arXiv:2307.13731 [hep-ph]} \BibitemShut
  {NoStop}%
\bibitem [{\citenamefont {De~la Torre~Luque}\ \emph
  {et~al.}(2023{\natexlab{c}})\citenamefont {De~la Torre~Luque}, \citenamefont
  {Balaji},\ and\ \citenamefont {Carenza}}]{DelaTorreLuque:2023nhh}%
  \BibitemOpen
  \bibfield  {author} {\bibinfo {author} {\bibfnamefont {P.}~\bibnamefont
  {De~la Torre~Luque}}, \bibinfo {author} {\bibfnamefont {S.}~\bibnamefont
  {Balaji}}, \ and\ \bibinfo {author} {\bibfnamefont {P.}~\bibnamefont
  {Carenza}},\ }\href@noop {} {\bibfield  {journal} {\bibinfo  {journal} {To
  appear in PRD as a letter}\ } (\bibinfo {year} {2023}{\natexlab{c}})},\
  \Eprint {http://arxiv.org/abs/2307.13728} {arXiv:2307.13728 [hep-ph]}
  \BibitemShut {NoStop}%
\bibitem [{\citenamefont {De~la Torre~Luque}\ \emph
  {et~al.}(2023{\natexlab{d}})\citenamefont {De~la Torre~Luque}, \citenamefont
  {Balaji},\ and\ \citenamefont {Koechler}}]{DelaTorreLuque:2023olp}%
  \BibitemOpen
  \bibfield  {author} {\bibinfo {author} {\bibfnamefont {P.}~\bibnamefont
  {De~la Torre~Luque}}, \bibinfo {author} {\bibfnamefont {S.}~\bibnamefont
  {Balaji}}, \ and\ \bibinfo {author} {\bibfnamefont {J.}~\bibnamefont
  {Koechler}},\ }\href@noop {} {\bibfield  {journal} {\bibinfo  {journal}
  {Accepted in The Astrophysical Journal}\ } (\bibinfo {year}
  {2023}{\natexlab{d}})},\ \Eprint {http://arxiv.org/abs/2311.04979}
  {arXiv:2311.04979 [hep-ph]} \BibitemShut {NoStop}%
\bibitem [{\citenamefont {{Ginzburg}}\ and\ \citenamefont
  {{Syrovatskii}}(1969)}]{Ginz&Syr}%
  \BibitemOpen
  \bibfield  {author} {\bibinfo {author} {\bibfnamefont {V.~L.}\ \bibnamefont
  {{Ginzburg}}}\ and\ \bibinfo {author} {\bibfnamefont {S.~I.}\ \bibnamefont
  {{Syrovatskii}}},\ }\href@noop {} {\emph {\bibinfo {title} {{The origin of
  cosmic rays}}}}\ (\bibinfo  {publisher} {Gordon \& Breach Publishing Group},\
  \bibinfo {year} {1969})\BibitemShut {NoStop}%
\bibitem [{\citenamefont {Crank}\ and\ \citenamefont
  {Nicolson}(1947)}]{Crank_Nicolson_1947}%
  \BibitemOpen
  \bibfield  {author} {\bibinfo {author} {\bibfnamefont {J.}~\bibnamefont
  {Crank}}\ and\ \bibinfo {author} {\bibfnamefont {P.}~\bibnamefont
  {Nicolson}},\ }\href {\doibase 10.1017/S0305004100023197} {\bibfield
  {journal} {\bibinfo  {journal} {Mathematical Proceedings of the Cambridge
  Philosophical Society}\ }\textbf {\bibinfo {volume} {43}},\ \bibinfo {pages}
  {50–67} (\bibinfo {year} {1947})}\BibitemShut {NoStop}%
\bibitem [{\citenamefont {Akiyama}\ \emph {et~al.}(2022)\citenamefont {Akiyama}
  \emph {et~al.}}]{EventHorizonTelescope:2022wkp}%
  \BibitemOpen
  \bibfield  {author} {\bibinfo {author} {\bibfnamefont {K.}~\bibnamefont
  {Akiyama}} \emph {et~al.} (\bibinfo {collaboration} {Event Horizon
  Telescope}),\ }\href {\doibase 10.3847/2041-8213/ac6674} {\bibfield
  {journal} {\bibinfo  {journal} {Astrophys. J. Lett.}\ }\textbf {\bibinfo
  {volume} {930}},\ \bibinfo {pages} {L12} (\bibinfo {year} {2022})},\ \Eprint
  {http://arxiv.org/abs/2311.08680} {arXiv:2311.08680 [astro-ph.HE]}
  \BibitemShut {NoStop}%
\bibitem [{\citenamefont {Aguilar}\ \emph {et~al.}(2021)\citenamefont {Aguilar}
  \emph {et~al.}}]{AGUILAR20211}%
  \BibitemOpen
  \bibfield  {author} {\bibinfo {author} {\bibfnamefont {M.}~\bibnamefont
  {Aguilar}} \emph {et~al.},\ }\href {\doibase
  https://doi.org/10.1016/j.physrep.2020.09.003} {\bibfield  {journal}
  {\bibinfo  {journal} {Physics Reports}\ }\textbf {\bibinfo {volume} {894}},\
  \bibinfo {pages} {1} (\bibinfo {year} {2021})},\ \bibinfo {note} {the Alpha
  Magnetic Spectrometer (AMS) on the International Space Station: Part II -
  Results from the First Seven Years}\BibitemShut {NoStop}%
\bibitem [{\citenamefont {Aguilar}\ \emph {et~al.}(2018)\citenamefont {Aguilar}
  \emph {et~al.}}]{aguilar2018observation}%
  \BibitemOpen
  \bibfield  {author} {\bibinfo {author} {\bibfnamefont {M.}~\bibnamefont
  {Aguilar}} \emph {et~al.} (\bibinfo {collaboration} {AMS Collaboration}),\
  }\href {\doibase 10.1103/PhysRevLett.120.021101} {\bibfield  {journal}
  {\bibinfo  {journal} {Phys. Rev. Lett.}\ }\textbf {\bibinfo {volume} {120}},\
  \bibinfo {pages} {021101} (\bibinfo {year} {2018})}\BibitemShut {NoStop}%
\bibitem [{\citenamefont {Aguilar}\ \emph {et~al.}(2013)\citenamefont {Aguilar}
  \emph {et~al.}}]{AMS_gen}%
  \BibitemOpen
  \bibfield  {author} {\bibinfo {author} {\bibfnamefont {M.}~\bibnamefont
  {Aguilar}} \emph {et~al.} (\bibinfo {collaboration} {AMS Collaboration}),\
  }\href {\doibase 10.1103/PhysRevLett.110.141102} {\bibfield  {journal}
  {\bibinfo  {journal} {Phys. Rev. Lett.}\ }\textbf {\bibinfo {volume} {110}},\
  \bibinfo {pages} {141102} (\bibinfo {year} {2013})}\BibitemShut {NoStop}%
\bibitem [{\citenamefont {Webber}\ \emph {et~al.}(2002)\citenamefont {Webber},
  \citenamefont {Lukasiak},\ and\ \citenamefont {Mcdonald}}]{VoyagerMO}%
  \BibitemOpen
  \bibfield  {author} {\bibinfo {author} {\bibfnamefont {W.}~\bibnamefont
  {Webber}}, \bibinfo {author} {\bibfnamefont {A.}~\bibnamefont {Lukasiak}}, \
  and\ \bibinfo {author} {\bibfnamefont {F.}~\bibnamefont {Mcdonald}},\
  }\href@noop {} {\bibfield  {journal} {\bibinfo  {journal} {ApJ}\ }\textbf
  {\bibinfo {volume} {568}},\ \bibinfo {pages} {210} (\bibinfo {year}
  {2002})}\BibitemShut {NoStop}%
\bibitem [{\citenamefont {Stone}\ \emph {et~al.}(2013)\citenamefont {Stone},
  \citenamefont {Cummings}, \citenamefont {McDonald}, \citenamefont {Heikkila},
  \citenamefont {Lal},\ and\ \citenamefont {Webber}}]{Stone150}%
  \BibitemOpen
  \bibfield  {author} {\bibinfo {author} {\bibfnamefont {E.~C.}\ \bibnamefont
  {Stone}}, \bibinfo {author} {\bibfnamefont {A.~C.}\ \bibnamefont {Cummings}},
  \bibinfo {author} {\bibfnamefont {F.~B.}\ \bibnamefont {McDonald}}, \bibinfo
  {author} {\bibfnamefont {B.~C.}\ \bibnamefont {Heikkila}}, \bibinfo {author}
  {\bibfnamefont {N.}~\bibnamefont {Lal}}, \ and\ \bibinfo {author}
  {\bibfnamefont {W.~R.}\ \bibnamefont {Webber}},\ }\href {\doibase
  10.1126/science.1236408} {\bibfield  {journal} {\bibinfo  {journal}
  {Science}\ }\textbf {\bibinfo {volume} {341}},\ \bibinfo {pages} {150}
  (\bibinfo {year} {2013})},\ \Eprint
  {http://arxiv.org/abs/https://science.sciencemag.org/content/341/6142/150.full.pdf}
  {https://science.sciencemag.org/content/341/6142/150.full.pdf} \BibitemShut
  {NoStop}%
\bibitem [{\citenamefont {Evoli}\ \emph {et~al.}(2020)\citenamefont {Evoli}
  \emph {et~al.}}]{Evoli_2020}%
  \BibitemOpen
  \bibfield  {author} {\bibinfo {author} {\bibfnamefont {C.}~\bibnamefont
  {Evoli}} \emph {et~al.},\ }\href {\doibase 10.1103/PhysRevLett.125.051101}
  {\bibfield  {journal} {\bibinfo  {journal} {Phys. Rev. Lett.}\ }\textbf
  {\bibinfo {volume} {125}},\ \bibinfo {pages} {051101} (\bibinfo {year}
  {2020})}\BibitemShut {NoStop}%
\bibitem [{\citenamefont {Cirelli}\ \emph {et~al.}(2011)\citenamefont
  {Cirelli}, \citenamefont {Corcella}, \citenamefont {Hektor}, \citenamefont
  {Hutsi}, \citenamefont {Kadastik}, \citenamefont {Panci}, \citenamefont
  {Raidal}, \citenamefont {Sala},\ and\ \citenamefont
  {Strumia}}]{Cirelli:2010xx}%
  \BibitemOpen
  \bibfield  {author} {\bibinfo {author} {\bibfnamefont {M.}~\bibnamefont
  {Cirelli}}, \bibinfo {author} {\bibfnamefont {G.}~\bibnamefont {Corcella}},
  \bibinfo {author} {\bibfnamefont {A.}~\bibnamefont {Hektor}}, \bibinfo
  {author} {\bibfnamefont {G.}~\bibnamefont {Hutsi}}, \bibinfo {author}
  {\bibfnamefont {M.}~\bibnamefont {Kadastik}}, \bibinfo {author}
  {\bibfnamefont {P.}~\bibnamefont {Panci}}, \bibinfo {author} {\bibfnamefont
  {M.}~\bibnamefont {Raidal}}, \bibinfo {author} {\bibfnamefont
  {F.}~\bibnamefont {Sala}}, \ and\ \bibinfo {author} {\bibfnamefont
  {A.}~\bibnamefont {Strumia}},\ }\href {\doibase
  10.1088/1475-7516/2012/10/E01} {\bibfield  {journal} {\bibinfo  {journal}
  {JCAP}\ }\textbf {\bibinfo {volume} {03}},\ \bibinfo {pages} {051} (\bibinfo
  {year} {2011})},\ \bibinfo {note} {[Erratum: JCAP 10, E01 (2012)]},\ \Eprint
  {http://arxiv.org/abs/1012.4515} {arXiv:1012.4515 [hep-ph]} \BibitemShut
  {NoStop}%
\bibitem [{\citenamefont {Boehm}\ \emph {et~al.}(2012)\citenamefont {Boehm},
  \citenamefont {Dolan},\ and\ \citenamefont {McCabe}}]{Boehm:2012gr}%
  \BibitemOpen
  \bibfield  {author} {\bibinfo {author} {\bibfnamefont {C.}~\bibnamefont
  {Boehm}}, \bibinfo {author} {\bibfnamefont {M.~J.}\ \bibnamefont {Dolan}}, \
  and\ \bibinfo {author} {\bibfnamefont {C.}~\bibnamefont {McCabe}},\ }\href
  {\doibase 10.1088/1475-7516/2012/12/027} {\bibfield  {journal} {\bibinfo
  {journal} {JCAP}\ }\textbf {\bibinfo {volume} {12}},\ \bibinfo {pages} {027}
  (\bibinfo {year} {2012})},\ \Eprint {http://arxiv.org/abs/1207.0497}
  {arXiv:1207.0497 [astro-ph.CO]} \BibitemShut {NoStop}%
\bibitem [{\citenamefont {Boehm}\ \emph {et~al.}(2013)\citenamefont {Boehm},
  \citenamefont {Dolan},\ and\ \citenamefont {McCabe}}]{Boehm:2013jpa}%
  \BibitemOpen
  \bibfield  {author} {\bibinfo {author} {\bibfnamefont {C.}~\bibnamefont
  {Boehm}}, \bibinfo {author} {\bibfnamefont {M.~J.}\ \bibnamefont {Dolan}}, \
  and\ \bibinfo {author} {\bibfnamefont {C.}~\bibnamefont {McCabe}},\ }\href
  {\doibase 10.1088/1475-7516/2013/08/041} {\bibfield  {journal} {\bibinfo
  {journal} {JCAP}\ }\textbf {\bibinfo {volume} {08}},\ \bibinfo {pages} {041}
  (\bibinfo {year} {2013})},\ \Eprint {http://arxiv.org/abs/1303.6270}
  {arXiv:1303.6270 [hep-ph]} \BibitemShut {NoStop}%
\bibitem [{\citenamefont {Nollett}\ and\ \citenamefont
  {Steigman}(2015)}]{Nollett:2014lwa}%
  \BibitemOpen
  \bibfield  {author} {\bibinfo {author} {\bibfnamefont {K.~M.}\ \bibnamefont
  {Nollett}}\ and\ \bibinfo {author} {\bibfnamefont {G.}~\bibnamefont
  {Steigman}},\ }\href {\doibase 10.1103/PhysRevD.91.083505} {\bibfield
  {journal} {\bibinfo  {journal} {Phys. Rev. D}\ }\textbf {\bibinfo {volume}
  {91}},\ \bibinfo {pages} {083505} (\bibinfo {year} {2015})},\ \Eprint
  {http://arxiv.org/abs/1411.6005} {arXiv:1411.6005 [astro-ph.CO]} \BibitemShut
  {NoStop}%
\bibitem [{\citenamefont {Escudero}(2019)}]{Escudero:2018mvt}%
  \BibitemOpen
  \bibfield  {author} {\bibinfo {author} {\bibfnamefont {M.}~\bibnamefont
  {Escudero}},\ }\href {\doibase 10.1088/1475-7516/2019/02/007} {\bibfield
  {journal} {\bibinfo  {journal} {JCAP}\ }\textbf {\bibinfo {volume} {02}},\
  \bibinfo {pages} {007} (\bibinfo {year} {2019})},\ \Eprint
  {http://arxiv.org/abs/1812.05605} {arXiv:1812.05605 [hep-ph]} \BibitemShut
  {NoStop}%
\bibitem [{\citenamefont {Sabti}\ \emph {et~al.}(2020)\citenamefont {Sabti},
  \citenamefont {Alvey}, \citenamefont {Escudero}, \citenamefont {Fairbairn},\
  and\ \citenamefont {Blas}}]{Sabti:2019mhn}%
  \BibitemOpen
  \bibfield  {author} {\bibinfo {author} {\bibfnamefont {N.}~\bibnamefont
  {Sabti}}, \bibinfo {author} {\bibfnamefont {J.}~\bibnamefont {Alvey}},
  \bibinfo {author} {\bibfnamefont {M.}~\bibnamefont {Escudero}}, \bibinfo
  {author} {\bibfnamefont {M.}~\bibnamefont {Fairbairn}}, \ and\ \bibinfo
  {author} {\bibfnamefont {D.}~\bibnamefont {Blas}},\ }\href {\doibase
  10.1088/1475-7516/2020/01/004} {\bibfield  {journal} {\bibinfo  {journal}
  {JCAP}\ }\textbf {\bibinfo {volume} {01}},\ \bibinfo {pages} {004} (\bibinfo
  {year} {2020})},\ \Eprint {http://arxiv.org/abs/1910.01649} {arXiv:1910.01649
  [hep-ph]} \BibitemShut {NoStop}%
\bibitem [{\citenamefont {Balan}\ \emph
  {et~al.}(2024{\natexlab{b}})\citenamefont {Balan}, \citenamefont {Balázs},
  \citenamefont {Bringmann}, \citenamefont {Cappiello}, \citenamefont {Catena},
  \citenamefont {Emken}, \citenamefont {Gonzalo}, \citenamefont {Gray},
  \citenamefont {Handley}, \citenamefont {Huynh}, \citenamefont {Kahlhoefer},\
  and\ \citenamefont {Vincent}}]{balan2024resonantasymmetricstatussubgev}%
  \BibitemOpen
  \bibfield  {author} {\bibinfo {author} {\bibfnamefont {S.}~\bibnamefont
  {Balan}}, \bibinfo {author} {\bibfnamefont {C.}~\bibnamefont {Balázs}},
  \bibinfo {author} {\bibfnamefont {T.}~\bibnamefont {Bringmann}}, \bibinfo
  {author} {\bibfnamefont {C.}~\bibnamefont {Cappiello}}, \bibinfo {author}
  {\bibfnamefont {R.}~\bibnamefont {Catena}}, \bibinfo {author} {\bibfnamefont
  {T.}~\bibnamefont {Emken}}, \bibinfo {author} {\bibfnamefont {T.~E.}\
  \bibnamefont {Gonzalo}}, \bibinfo {author} {\bibfnamefont {T.~R.}\
  \bibnamefont {Gray}}, \bibinfo {author} {\bibfnamefont {W.}~\bibnamefont
  {Handley}}, \bibinfo {author} {\bibfnamefont {Q.}~\bibnamefont {Huynh}},
  \bibinfo {author} {\bibfnamefont {F.}~\bibnamefont {Kahlhoefer}}, \ and\
  \bibinfo {author} {\bibfnamefont {A.~C.}\ \bibnamefont {Vincent}},\ }\href
  {https://arxiv.org/abs/2405.17548} {\enquote {\bibinfo {title} {Resonant or
  asymmetric: The status of sub-gev dark matter},}\ } (\bibinfo {year}
  {2024}{\natexlab{b}}),\ \Eprint {http://arxiv.org/abs/2405.17548}
  {arXiv:2405.17548 [hep-ph]} \BibitemShut {NoStop}%
\bibitem [{\citenamefont {Moore}\ \emph {et~al.}(1999)\citenamefont {Moore}
  \emph {et~al.}}]{Moore_1999}%
  \BibitemOpen
  \bibfield  {author} {\bibinfo {author} {\bibfnamefont {B.}~\bibnamefont
  {Moore}} \emph {et~al.},\ }\href {\doibase 10.1086/312287} {\bibfield
  {journal} {\bibinfo  {journal} {The Astrophysical Journal}\ }\textbf
  {\bibinfo {volume} {524}},\ \bibinfo {pages} {L19–L22} (\bibinfo {year}
  {1999})}\BibitemShut {NoStop}%
\bibitem [{\citenamefont {Ackermann}\ \emph {et~al.}(2017)\citenamefont
  {Ackermann} \emph {et~al.}}]{Ackermann_2017}%
  \BibitemOpen
  \bibfield  {author} {\bibinfo {author} {\bibfnamefont {M.}~\bibnamefont
  {Ackermann}} \emph {et~al.},\ }\href {\doibase 10.3847/1538-4357/aa6cab}
  {\bibfield  {journal} {\bibinfo  {journal} {The Astrophysical Journal}\
  }\textbf {\bibinfo {volume} {840}},\ \bibinfo {pages} {43} (\bibinfo {year}
  {2017})}\BibitemShut {NoStop}%
\bibitem [{\citenamefont {Di~Mauro}(2021)}]{Di_Mauro_2021}%
  \BibitemOpen
  \bibfield  {author} {\bibinfo {author} {\bibfnamefont {M.}~\bibnamefont
  {Di~Mauro}},\ }\href {\doibase 10.1103/physrevd.103.063029} {\bibfield
  {journal} {\bibinfo  {journal} {Physical Review D}\ }\textbf {\bibinfo
  {volume} {103}} (\bibinfo {year} {2021}),\
  10.1103/physrevd.103.063029}\BibitemShut {NoStop}%
\bibitem [{\citenamefont {Seo}\ and\ \citenamefont
  {Ptuskin}(1994)}]{seo1994stochastic}%
  \BibitemOpen
  \bibfield  {author} {\bibinfo {author} {\bibfnamefont {E.-S.}\ \bibnamefont
  {Seo}}\ and\ \bibinfo {author} {\bibfnamefont {V.~S.}\ \bibnamefont
  {Ptuskin}},\ }\href@noop {} {\bibfield  {journal} {\bibinfo  {journal} {ApJ}\
  }\textbf {\bibinfo {volume} {431}},\ \bibinfo {pages} {705} (\bibinfo {year}
  {1994})}\BibitemShut {NoStop}%
\bibitem [{\citenamefont {Dundovic}\ \emph {et~al.}(2021)\citenamefont
  {Dundovic} \emph {et~al.}}]{Dundovic:2021ryb}%
  \BibitemOpen
  \bibfield  {author} {\bibinfo {author} {\bibfnamefont {A.}~\bibnamefont
  {Dundovic}} \emph {et~al.},\ }\href {\doibase 10.1051/0004-6361/202140801}
  {\bibfield  {journal} {\bibinfo  {journal} {Astron. Astrophys.}\ }\textbf
  {\bibinfo {volume} {653}},\ \bibinfo {pages} {A18} (\bibinfo {year}
  {2021})},\ \Eprint {http://arxiv.org/abs/2105.13165} {arXiv:2105.13165}
  \BibitemShut {NoStop}%
\bibitem [{\citenamefont {Foster}\ \emph {et~al.}(2021)\citenamefont {Foster},
  \citenamefont {Kongsore}, \citenamefont {Dessert}, \citenamefont {Park},
  \citenamefont {Rodd}, \citenamefont {Cranmer},\ and\ \citenamefont
  {Safdi}}]{XMM}%
  \BibitemOpen
  \bibfield  {author} {\bibinfo {author} {\bibfnamefont {J.~W.}\ \bibnamefont
  {Foster}}, \bibinfo {author} {\bibfnamefont {M.}~\bibnamefont {Kongsore}},
  \bibinfo {author} {\bibfnamefont {C.}~\bibnamefont {Dessert}}, \bibinfo
  {author} {\bibfnamefont {Y.}~\bibnamefont {Park}}, \bibinfo {author}
  {\bibfnamefont {N.~L.}\ \bibnamefont {Rodd}}, \bibinfo {author}
  {\bibfnamefont {K.}~\bibnamefont {Cranmer}}, \ and\ \bibinfo {author}
  {\bibfnamefont {B.~R.}\ \bibnamefont {Safdi}},\ }\href {\doibase
  10.1103/PhysRevLett.127.051101} {\bibfield  {journal} {\bibinfo  {journal}
  {Phys. Rev. Lett.}\ }\textbf {\bibinfo {volume} {127}},\ \bibinfo {pages}
  {051101} (\bibinfo {year} {2021})}\BibitemShut {NoStop}%
\bibitem [{\citenamefont {Cirelli}\ \emph
  {et~al.}(2023{\natexlab{b}})\citenamefont {Cirelli}, \citenamefont
  {Fornengo}, \citenamefont {Koechler}, \citenamefont {Pinetti},\ and\
  \citenamefont {Roach}}]{Cirelli_2023}%
  \BibitemOpen
  \bibfield  {author} {\bibinfo {author} {\bibfnamefont {M.}~\bibnamefont
  {Cirelli}}, \bibinfo {author} {\bibfnamefont {N.}~\bibnamefont {Fornengo}},
  \bibinfo {author} {\bibfnamefont {J.}~\bibnamefont {Koechler}}, \bibinfo
  {author} {\bibfnamefont {E.}~\bibnamefont {Pinetti}}, \ and\ \bibinfo
  {author} {\bibfnamefont {B.~M.}\ \bibnamefont {Roach}},\ }\href {\doibase
  10.1088/1475-7516/2023/07/026} {\bibfield  {journal} {\bibinfo  {journal}
  {Journal of Cosmology and Astroparticle Physics}\ }\textbf {\bibinfo {volume}
  {2023}},\ \bibinfo {pages} {026} (\bibinfo {year}
  {2023}{\natexlab{b}})}\BibitemShut {NoStop}%
\bibitem [{\citenamefont {Luque}\ \emph {et~al.}(2024)\citenamefont {Luque},
  \citenamefont {Balaji},\ and\ \citenamefont {Carenza}}]{Luque_2024}%
  \BibitemOpen
  \bibfield  {author} {\bibinfo {author} {\bibfnamefont {P.~D. l.~T.}\
  \bibnamefont {Luque}}, \bibinfo {author} {\bibfnamefont {S.}~\bibnamefont
  {Balaji}}, \ and\ \bibinfo {author} {\bibfnamefont {P.}~\bibnamefont
  {Carenza}},\ }\href {\doibase 10.1103/physrevd.109.103028} {\bibfield
  {journal} {\bibinfo  {journal} {Physical Review D}\ }\textbf {\bibinfo
  {volume} {109}} (\bibinfo {year} {2024}),\
  10.1103/physrevd.109.103028}\BibitemShut {NoStop}%
\bibitem [{\citenamefont {Cirelli}\ \emph {et~al.}(2013)\citenamefont
  {Cirelli}, \citenamefont {Serpico},\ and\ \citenamefont
  {Zaharijas}}]{Cirelli_Bremss}%
  \BibitemOpen
  \bibfield  {author} {\bibinfo {author} {\bibfnamefont {M.}~\bibnamefont
  {Cirelli}}, \bibinfo {author} {\bibfnamefont {P.~D.}\ \bibnamefont
  {Serpico}}, \ and\ \bibinfo {author} {\bibfnamefont {G.}~\bibnamefont
  {Zaharijas}},\ }\href {\doibase 10.1088/1475-7516/2013/11/035} {\bibfield
  {journal} {\bibinfo  {journal} {Journal of Cosmology and Astroparticle
  Physics}\ }\textbf {\bibinfo {volume} {2013}},\ \bibinfo {pages} {035}
  (\bibinfo {year} {2013})}\BibitemShut {NoStop}%
\bibitem [{\citenamefont {Bartels}\ \emph {et~al.}(2017)\citenamefont
  {Bartels}, \citenamefont {Gaggero},\ and\ \citenamefont
  {Weniger}}]{Bartels_2017}%
  \BibitemOpen
  \bibfield  {author} {\bibinfo {author} {\bibfnamefont {R.}~\bibnamefont
  {Bartels}}, \bibinfo {author} {\bibfnamefont {D.}~\bibnamefont {Gaggero}}, \
  and\ \bibinfo {author} {\bibfnamefont {C.}~\bibnamefont {Weniger}},\ }\href
  {\doibase 10.1088/1475-7516/2017/05/001} {\bibfield  {journal} {\bibinfo
  {journal} {Journal of Cosmology and Astroparticle Physics}\ }\textbf
  {\bibinfo {volume} {2017}},\ \bibinfo {pages} {001–001} (\bibinfo {year}
  {2017})}\BibitemShut {NoStop}%
\bibitem [{\citenamefont {Bouchet}\ \emph {et~al.}(2010)\citenamefont
  {Bouchet}, \citenamefont {Roques},\ and\ \citenamefont
  {Jourdain}}]{Bouchet:2010dj}%
  \BibitemOpen
  \bibfield  {author} {\bibinfo {author} {\bibfnamefont {L.}~\bibnamefont
  {Bouchet}}, \bibinfo {author} {\bibfnamefont {J.-P.}\ \bibnamefont {Roques}},
  \ and\ \bibinfo {author} {\bibfnamefont {E.}~\bibnamefont {Jourdain}},\
  }\href {\doibase 10.1088/0004-637X/720/2/1772} {\bibfield  {journal}
  {\bibinfo  {journal} {Astrophys. J.}\ }\textbf {\bibinfo {volume} {720}},\
  \bibinfo {pages} {1772} (\bibinfo {year} {2010})},\ \Eprint
  {http://arxiv.org/abs/1007.4753} {arXiv:1007.4753 [astro-ph.HE]} \BibitemShut
  {NoStop}%
\bibitem [{\citenamefont {Karwin}\ \emph {et~al.}(2023)\citenamefont {Karwin},
  \citenamefont {Siegert}, \citenamefont {Beechert}, \citenamefont {Tomsick},
  \citenamefont {Porter}, \citenamefont {Negro}, \citenamefont {Kierans},
  \citenamefont {Ajello}, \citenamefont {Martinez-Castellanos}, \citenamefont
  {Shih}, \citenamefont {Zoglauer},\ and\ \citenamefont {Boggs}}]{Karwin_2023}%
  \BibitemOpen
  \bibfield  {author} {\bibinfo {author} {\bibfnamefont {C.~M.}\ \bibnamefont
  {Karwin}}, \bibinfo {author} {\bibfnamefont {T.}~\bibnamefont {Siegert}},
  \bibinfo {author} {\bibfnamefont {J.}~\bibnamefont {Beechert}}, \bibinfo
  {author} {\bibfnamefont {J.~A.}\ \bibnamefont {Tomsick}}, \bibinfo {author}
  {\bibfnamefont {T.~A.}\ \bibnamefont {Porter}}, \bibinfo {author}
  {\bibfnamefont {M.}~\bibnamefont {Negro}}, \bibinfo {author} {\bibfnamefont
  {C.}~\bibnamefont {Kierans}}, \bibinfo {author} {\bibfnamefont
  {M.}~\bibnamefont {Ajello}}, \bibinfo {author} {\bibfnamefont
  {I.}~\bibnamefont {Martinez-Castellanos}}, \bibinfo {author} {\bibfnamefont
  {A.}~\bibnamefont {Shih}}, \bibinfo {author} {\bibfnamefont {A.}~\bibnamefont
  {Zoglauer}}, \ and\ \bibinfo {author} {\bibfnamefont {S.~E.}\ \bibnamefont
  {Boggs}},\ }\href {\doibase 10.3847/1538-4357/ad04df} {\bibfield  {journal}
  {\bibinfo  {journal} {The Astrophysical Journal}\ }\textbf {\bibinfo {volume}
  {959}},\ \bibinfo {pages} {90} (\bibinfo {year} {2023})}\BibitemShut
  {NoStop}%
\bibitem [{\citenamefont {Beacom}\ and\ \citenamefont
  {Yuksel}(2006)}]{Beacom:2005qv}%
  \BibitemOpen
  \bibfield  {author} {\bibinfo {author} {\bibfnamefont {J.~F.}\ \bibnamefont
  {Beacom}}\ and\ \bibinfo {author} {\bibfnamefont {H.}~\bibnamefont
  {Yuksel}},\ }\href {\doibase 10.1103/PhysRevLett.97.071102} {\bibfield
  {journal} {\bibinfo  {journal} {Phys. Rev. Lett.}\ }\textbf {\bibinfo
  {volume} {97}},\ \bibinfo {pages} {071102} (\bibinfo {year} {2006})},\
  \Eprint {http://arxiv.org/abs/astro-ph/0512411} {arXiv:astro-ph/0512411}
  \BibitemShut {NoStop}%
\bibitem [{\citenamefont {Sizun}\ \emph {et~al.}(2006)\citenamefont {Sizun},
  \citenamefont {Casse},\ and\ \citenamefont {Schanne}}]{Sizun:2006uh}%
  \BibitemOpen
  \bibfield  {author} {\bibinfo {author} {\bibfnamefont {P.}~\bibnamefont
  {Sizun}}, \bibinfo {author} {\bibfnamefont {M.}~\bibnamefont {Casse}}, \ and\
  \bibinfo {author} {\bibfnamefont {S.}~\bibnamefont {Schanne}},\ }\href
  {\doibase 10.1103/PhysRevD.74.063514} {\bibfield  {journal} {\bibinfo
  {journal} {Phys. Rev. D}\ }\textbf {\bibinfo {volume} {74}},\ \bibinfo
  {pages} {063514} (\bibinfo {year} {2006})},\ \Eprint
  {http://arxiv.org/abs/astro-ph/0607374} {arXiv:astro-ph/0607374} \BibitemShut
  {NoStop}%
\bibitem [{\citenamefont {Sizun}\ \emph {et~al.}(2007)\citenamefont {Sizun},
  \citenamefont {Casse}, \citenamefont {Schanne},\ and\ \citenamefont
  {Cordier}}]{Sizun:2007ds}%
  \BibitemOpen
  \bibfield  {author} {\bibinfo {author} {\bibfnamefont {P.}~\bibnamefont
  {Sizun}}, \bibinfo {author} {\bibfnamefont {M.}~\bibnamefont {Casse}},
  \bibinfo {author} {\bibfnamefont {S.}~\bibnamefont {Schanne}}, \ and\
  \bibinfo {author} {\bibfnamefont {B.}~\bibnamefont {Cordier}},\ }\href@noop
  {} {\bibfield  {journal} {\bibinfo  {journal} {ESA Spec. Publ.}\ }\textbf
  {\bibinfo {volume} {622}},\ \bibinfo {pages} {61} (\bibinfo {year} {2007})},\
  \Eprint {http://arxiv.org/abs/astro-ph/0702061} {arXiv:astro-ph/0702061}
  \BibitemShut {NoStop}%
\bibitem [{\citenamefont {De~la Torre~Luque}\ \emph
  {et~al.}(2024{\natexlab{a}})\citenamefont {De~la Torre~Luque}, \citenamefont
  {Balaji}, \citenamefont {Fairbairn}, \citenamefont {Sala},\ and\
  \citenamefont {Silk}}]{DelaTorreLuque:2024wfz}%
  \BibitemOpen
  \bibfield  {author} {\bibinfo {author} {\bibfnamefont {P.}~\bibnamefont
  {De~la Torre~Luque}}, \bibinfo {author} {\bibfnamefont {S.}~\bibnamefont
  {Balaji}}, \bibinfo {author} {\bibfnamefont {M.}~\bibnamefont {Fairbairn}},
  \bibinfo {author} {\bibfnamefont {F.}~\bibnamefont {Sala}}, \ and\ \bibinfo
  {author} {\bibfnamefont {J.}~\bibnamefont {Silk}},\ }\href@noop {} {\
  (\bibinfo {year} {2024}{\natexlab{a}})},\ \Eprint
  {http://arxiv.org/abs/2410.16379} {arXiv:2410.16379 [astro-ph.HE]}
  \BibitemShut {NoStop}%
\bibitem [{\citenamefont {De~la Torre~Luque}\ \emph
  {et~al.}(2024{\natexlab{b}})\citenamefont {De~la Torre~Luque}, \citenamefont
  {Balaji}, \citenamefont {Carenza},\ and\ \citenamefont
  {Mastrototaro}}]{DelaTorreLuque:2024zsr}%
  \BibitemOpen
  \bibfield  {author} {\bibinfo {author} {\bibfnamefont {P.}~\bibnamefont
  {De~la Torre~Luque}}, \bibinfo {author} {\bibfnamefont {S.}~\bibnamefont
  {Balaji}}, \bibinfo {author} {\bibfnamefont {P.}~\bibnamefont {Carenza}}, \
  and\ \bibinfo {author} {\bibfnamefont {L.}~\bibnamefont {Mastrototaro}},\
  }\href@noop {} {\  (\bibinfo {year} {2024}{\natexlab{b}})},\ \Eprint
  {http://arxiv.org/abs/2405.08482} {arXiv:2405.08482 [hep-ph]} \BibitemShut
  {NoStop}%
\bibitem [{\citenamefont {Calore}\ \emph {et~al.}(2021)\citenamefont {Calore},
  \citenamefont {Carenza}, \citenamefont {Giannotti}, \citenamefont {Jaeckel},
  \citenamefont {Lucente},\ and\ \citenamefont {Mirizzi}}]{Calore:2021klc}%
  \BibitemOpen
  \bibfield  {author} {\bibinfo {author} {\bibfnamefont {F.}~\bibnamefont
  {Calore}}, \bibinfo {author} {\bibfnamefont {P.}~\bibnamefont {Carenza}},
  \bibinfo {author} {\bibfnamefont {M.}~\bibnamefont {Giannotti}}, \bibinfo
  {author} {\bibfnamefont {J.}~\bibnamefont {Jaeckel}}, \bibinfo {author}
  {\bibfnamefont {G.}~\bibnamefont {Lucente}}, \ and\ \bibinfo {author}
  {\bibfnamefont {A.}~\bibnamefont {Mirizzi}},\ }\href {\doibase
  10.1103/PhysRevD.104.043016} {\bibfield  {journal} {\bibinfo  {journal}
  {Phys. Rev. D}\ }\textbf {\bibinfo {volume} {104}},\ \bibinfo {pages}
  {043016} (\bibinfo {year} {2021})},\ \Eprint
  {http://arxiv.org/abs/2107.02186} {arXiv:2107.02186 [hep-ph]} \BibitemShut
  {NoStop}%
\bibitem [{\citenamefont {Calore}\ \emph {et~al.}(2022)\citenamefont {Calore},
  \citenamefont {Carenza}, \citenamefont {Giannotti}, \citenamefont {Jaeckel},
  \citenamefont {Lucente}, \citenamefont {Mastrototaro},\ and\ \citenamefont
  {Mirizzi}}]{Calore:2021lih}%
  \BibitemOpen
  \bibfield  {author} {\bibinfo {author} {\bibfnamefont {F.}~\bibnamefont
  {Calore}}, \bibinfo {author} {\bibfnamefont {P.}~\bibnamefont {Carenza}},
  \bibinfo {author} {\bibfnamefont {M.}~\bibnamefont {Giannotti}}, \bibinfo
  {author} {\bibfnamefont {J.}~\bibnamefont {Jaeckel}}, \bibinfo {author}
  {\bibfnamefont {G.}~\bibnamefont {Lucente}}, \bibinfo {author} {\bibfnamefont
  {L.}~\bibnamefont {Mastrototaro}}, \ and\ \bibinfo {author} {\bibfnamefont
  {A.}~\bibnamefont {Mirizzi}},\ }\href {\doibase 10.1103/PhysRevD.105.063026}
  {\bibfield  {journal} {\bibinfo  {journal} {Phys. Rev. D}\ }\textbf {\bibinfo
  {volume} {105}},\ \bibinfo {pages} {063026} (\bibinfo {year} {2022})},\
  \Eprint {http://arxiv.org/abs/2112.08382} {arXiv:2112.08382 [hep-ph]}
  \BibitemShut {NoStop}%
\bibitem [{\citenamefont {Carenza}\ \emph {et~al.}(2024)\citenamefont
  {Carenza}, \citenamefont {Lucente}, \citenamefont {Mastrototaro},
  \citenamefont {Mirizzi},\ and\ \citenamefont {Serpico}}]{Carenza:2023old}%
  \BibitemOpen
  \bibfield  {author} {\bibinfo {author} {\bibfnamefont {P.}~\bibnamefont
  {Carenza}}, \bibinfo {author} {\bibfnamefont {G.}~\bibnamefont {Lucente}},
  \bibinfo {author} {\bibfnamefont {L.}~\bibnamefont {Mastrototaro}}, \bibinfo
  {author} {\bibfnamefont {A.}~\bibnamefont {Mirizzi}}, \ and\ \bibinfo
  {author} {\bibfnamefont {P.~D.}\ \bibnamefont {Serpico}},\ }\href {\doibase
  10.1103/PhysRevD.109.063010} {\bibfield  {journal} {\bibinfo  {journal}
  {Phys. Rev. D}\ }\textbf {\bibinfo {volume} {109}},\ \bibinfo {pages}
  {063010} (\bibinfo {year} {2024})},\ \Eprint
  {http://arxiv.org/abs/2311.00033} {arXiv:2311.00033 [hep-ph]} \BibitemShut
  {NoStop}%
\bibitem [{\citenamefont {{Guessoum}}\ \emph {et~al.}(1991)\citenamefont
  {{Guessoum}}, \citenamefont {{Ramaty}},\ and\ \citenamefont
  {{Lingenfelter}}}]{Guessoum1991}%
  \BibitemOpen
  \bibfield  {author} {\bibinfo {author} {\bibfnamefont {N.}~\bibnamefont
  {{Guessoum}}}, \bibinfo {author} {\bibfnamefont {R.}~\bibnamefont
  {{Ramaty}}}, \ and\ \bibinfo {author} {\bibfnamefont {R.~E.}\ \bibnamefont
  {{Lingenfelter}}},\ }\href {\doibase 10.1086/170417} {\bibfield  {journal}
  {\bibinfo  {journal} {The astrophysical Journal}\ }\textbf {\bibinfo {volume}
  {378}},\ \bibinfo {pages} {170} (\bibinfo {year} {1991})}\BibitemShut
  {NoStop}%
\bibitem [{\citenamefont {Siegert}\ \emph {et~al.}(2020)\citenamefont {Siegert}
  \emph {et~al.}}]{Siegert:2020oxw}%
  \BibitemOpen
  \bibfield  {author} {\bibinfo {author} {\bibfnamefont {T.}~\bibnamefont
  {Siegert}} \emph {et~al.},\ }\href {\doibase 10.3847/1538-4357/ab9607}
  {\bibfield  {journal} {\bibinfo  {journal} {Astrophys. J.}\ }\textbf
  {\bibinfo {volume} {897}},\ \bibinfo {pages} {45} (\bibinfo {year} {2020})},\
  \Eprint {http://arxiv.org/abs/2005.10950} {arXiv:2005.10950 [astro-ph.HE]}
  \BibitemShut {NoStop}%
\end{thebibliography}%

\end{document}